\documentclass{JHEP3}

\usepackage{epsfig}
\usepackage{amsmath}
\usepackage{color}
\usepackage{colortbl}

\newcommand{\slR}{{\tilde{l}_{\rm R}}}

\newcommand{\neu}{\tilde{\chi}^0}

\newcommand{\mneu}[1]{m_{\tilde{\chi}^0_{#1}}}

\newcommand{\one}{{(1,0)}}

\def\mathswitch#1{\relax\ifmmode#1\else$#1$\fi}
\def\mathswitchr#1{\relax\ifmmode{\mathrm{#1}}\else$\mathrm{#1}$\fi}

\newcommand{\SLASH}[2]{\makebox[#2ex][l]{$#1$}/}

\newcommand{\pslash}{\SLASH{p}{.2}}

\newcommand{\ETmiss}{{{\not{\!\!E}}_\top}}

\newcommand{\lesim}{\,\raisebox{-.1ex}{$_{\textstyle
<}\atop^{\textstyle\sim}$}\,}

\renewcommand{\figurename}{\bf Figure}
\renewcommand{\tablename}{\bf Table}
\newcommand{\mycaption}[1]{\caption{\sl #1}}

\definecolor{mymagenta}{rgb}{0.5,0,0.5}
\definecolor{mygreen}{rgb}{0,0.67,0}
\definecolor{lred}{rgb}{1,0.85,0.85}
\definecolor{lblue}{rgb}{0.85,0.85,1}
\definecolor{grey}{rgb}{0.9,0.9,0.9}

\title{Distinguishing spins in decay chains with photons at the Large Hadron Collider}

\author{Wolfgang~Ehrenfeld\\
  DESY, Hamburg, Notkestr. 85, D-22603 Hamburg, Germany\\
  E-mail: \email{wolfgang.ehrenfeld@desy.de}}
\author{Ayres~Freitas\\
  Department of Physics \& Astronomy, University of Pittsburgh,
  3941 O'Hara St, Pittsburgh, PA 15260, USA  \\
  E-mail: \email{afreitas@pitt.edu}}
\author{Ananda~Landwehr\thanks{Current address: Max-Planck-Institut f{\"u}r Physik,
    F\"ohringer Ring 6, D-80805 M{\"u}nchen, Germany}, Daniel~Wyler \\
  Institut f\"ur Theoretische Physik,
  Universit\"at Z\"urich, Winterthurerstrasse 190, 
  CH-8057 Z\"urich, Switzerland \\
  E-mail: \email{landwehr@mppmu.mpg.de}, \email{wyler@physik.uzh.ch}}

\preprint{\arXivid{0904.1293}}      

\abstract{
Several models for physics beyond the Standard Model predict new particles with
a decay signature including hard photons and missing energy. Two well-motivated
examples are supersymmetry with 
gauge-mediated breaking (GMSB) 
and the standard model with two universal extra
dimensions. 
Both models lead to decay chains with similar collider signatures,
including hard photon emission.
The main discriminating feature are the spins of the new particles.
In this paper we discuss how information about the spins of the particles
can be extracted from lepton-photon or quark-photon invariant mass distributions
at the Large Hadron Collider. The characteristic shapes of the distributions are
derived analytically and then studied in a realistic Monte-Carlo simulation. We
find that for a typical GMSB mass spectrum with particle masses below 1~TeV,
already 10~fb$^{-1}$ integrated luminosity at 14~TeV center-of-mass
energy are sufficient to discriminate the two
models with high significance.
}


\begin{document}


\section{Introduction}

Several models for physics beyond the Standard Model (SM) introduce partner
particles for all SM particles, the lightest of which is stable and could be
the constituent of dark matter.  The best known examples are supersymmetry
(SUSY) with conserved R-parity and universal extra dimensions (UED) with
conserved Kaluza-Klein (KK) parity \cite{Appelquist:2000nn}. Since the observable  signatures
for these models at the Large Hadron Collider (LHC) look quite similar
\cite{Cheng:2002ab}, it will be important to test the fundamental quantum
numbers of the new particles in order to scrutinize the nature of the
underlying physics.

For example, both SUSY and UED require that the couplings of the new particles
are identical to the corresponding couplings of their SM partners; a prediction
which can be tested at the LHC by measuring cross section ratios
\cite{Freitas:2007fd}. However, a crucial distinction between the two models is
given by the spins of the new particles. While the SUSY partners differ from
their SM counterparts by one half-unit of spin, the KK excitations in UED have
the same spin as their SM partners. Recently, extensive work has been performed
to determine the spins of SUSY or UED particles by exploiting angular
correlations in the decay of those particles at the LHC\footnote{A few studies
have explored more model-dependent discrimination methods based on total cross
sections \cite{Datta:2005vx}  and higher KK modes \cite{KK2}.}. Many papers have
focused on decay chains involving lepton pairs
\cite{ll,KK2,Miller:2005zp,ll2,Wang:2006hk,kwy,ll3,BKMP}. A typical example of such a
decay chain in SUSY is $\tilde{q} \to q \, \neu_2 \to q \, l^+ l^- \, \neu_1$.
Other studies have examined channels involving heavy gauge bosons
\cite{Wang:2006hk,Smillie:2006cd,Buckley:2007th,Buckley:2008pp}, sleptons
\cite{Buckley:2007th,slepton} and top quarks. However, in all of the existing
studies it was assumed that the lightest new particle is either a neutralino or weak
vector boson in SUSY or UED, respectively.

In this article the spin determination from angular correlations is extended to
decay chains that involve hard photons in the final state. Such decay channels
occur naturally in gauge-mediated supersymmetry breaking (GMSB), where the
lightest SUSY particle is the gravitino \cite{Dine:1994vc}, as well as in
the extension of the SM by two universal extra dimensions (UED6), where the lightest KK mode is
typically a scalar component of a higher-dimensional vector boson, called
``scalar adjoint'' \cite{ued6,6dsm,uedlhc}. It has been shown earlier that a
high-energy $e^+e^-$ collider could distinguish between GMSB and UED6 by
studying angular correlations in pair production and decay of the selectron
(KK-electrons) and neutralinos (KK-gauge bosons) \cite{Freitas:2007rh}. The
purpose of the present paper is to study how such a distinction can be achieved
at the LHC by analyzing decay chains involving leptons and photons. In
particular, we are investigating how the spin of the lightest new particle,
which escapes from the detector in form of missing momentum, can be inferred
from invariant mass distributions of the leptons and photons.

In section~\ref{sc:spinc} we describe analytical calculations of the relevant
invariant mass distributions and compare the angular correlations predicted by
GMSB and UED6. In order to evaluate the prospects for  experimental
measurements of these distributions, we present in section~\ref{sc:mc} results
of a realistic Monte-Carlo simulation, incorporating the spin correlation
effects of the two models. We present our conclusions in
section~\ref{sc:concl}. The detailed analytical results are collected in the
appendix.


\section{Spin correlations in GMSB and UED6}
\label{sc:spinc}

In our notations and conventions we follow Ref.~\cite{Martin:1997ns} for
supersymmetry and
Ref.~\cite{ued6,6dsm,uedlhc} for the Standard Model in six dimensions (UED6).

GMSB is a promising candidate for a
mechanism that generates TeV-scale masses for the SUSY partners,
most notably since it explains the absence of large flavor-changing neutral
currents. In GMSB, the lightest SUSY particle is typically the gravitino
$\tilde{G}$, with a mass $m_{\tilde{G}} \lesim 1$~MeV. If the next-to-lightest
SUSY particle is a neutralino, it will mostly decay into the gravitino and a
photon, $\neu_1 \to \gamma \, \tilde{G}$. Depending on the neutralino mass,
there could be a smaller branching fraction into a $Z$ boson, which we will not
investigate further.

In UED6, the lightest particle at KK level (1,0) is typically the scalar adjoint
of the hypercharge boson, $B_H^\one$ \cite{6dsm,Ponton:2005kx}. The vector
mode of the KK-hypercharge boson, $B_\mu^\one$, can decay into the scalar adjoint
via a loop-induced process, $B_\mu^\one \to \gamma \, B_H^\one$. This decay mode
has a sizable branching fraction of about 34\% \cite{uedlhc}.
Therefore the two models lead to very similar decay signatures. 
In particular, the decay chain
\begin{align}
&\neu_2 \to l^\pm \, \slR^\mp \to l^+ l^- \, \neu_1 \to
l^+ l^- \, \gamma \, \tilde{G}\,,
\label{eq:longch1}
\intertext{%
which is typical in GMSB,
is imitated by the equivalent process in UED6,}
&Z_\mu^\one \to l^\pm \, L_+^\one \to l^+ l^- \, B_\mu^\one \to
l^+ l^- \, \gamma \, B_H^\one\,,
\label{eq:longch2}
\end{align}
see Fig.~\ref{fig:chain1}.
Both processes lead to a final state signature of a same-flavor, opposite-sign
lepton pair, one photon, and missing transverse momentum $\ETmiss$.
\FIGURE[ht]{
\centering
\epsfig{figure=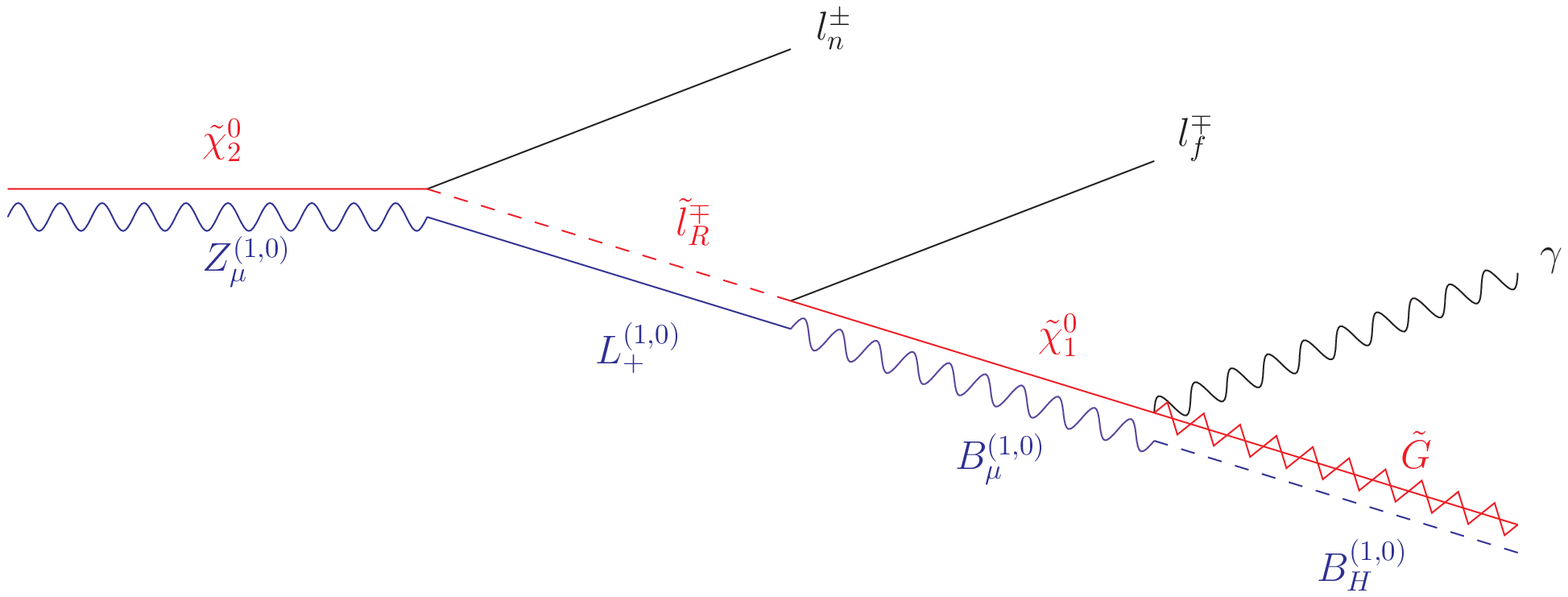, width=11cm}
\mycaption{SUSY (red/top) and UED6 (blue/bottom) decay chains with observable
final state $l^+l^-\gamma+{\ETmiss}$.}\label{fig:chain1}
}


In the following we will study spin correlation effects in these decay chains in
detail. We will also briefly analyze the following shorter decay modes of
squarks:\begin{align}
&\tilde{q}_R \to q \, \neu_1 \to q \, \gamma \, \tilde{G}
\label{eq:shortch1}
\intertext{%
for GMSB, and}
&Q_-^\one \to q \, B_\mu^\one \to q \, \gamma \, B_H^\one
\label{eq:shortch2}
\end{align}
for UED6. Although these decay channels have considerably larger SM backgrounds
at the LHC, they also have larger branching ratios compared to the decay channels
with leptons. They could be very useful to obtain information on the spin of the
quark partners.

While both GMSB and UED6 have similar mass hierarchies, which allow the decay
channels in eqs.~\eqref{eq:longch1}--\eqref{eq:shortch2}, the typical mass
spectra are quantitatively rather different.


In minimal gauge mediation the gaugino mass parameter relation $M_1/g_1^2
=M_2/g_2^2 = M_3/g_3^2$ implies that the weak gauginos are rather light, while
the gluino is much heavier. Furthermore, the squarks are also heavy, while the
sleptons have masses close the the gauginos. For our numerical analysis in the
next section we will use the reference scenario G1a from Ref.~\cite{hp}.
The masses of the particles appearing in our decay chain and their branching
ratios are summarized in Tab.~\ref{tab:scen}.
\TABLE[ht]{
  \centering
  \renewcommand{\arraystretch}{1.1}
  \begin{tabular}{|ll|ll|}
    \hline
    \multicolumn{2}{|c|}{G1a} &
    \multicolumn{2}{|c|}{U1} \\
    \hline
    Particle & Mass [GeV] & Particle & Mass [GeV]  \\
    \hline
    $\quad\tilde{g}$ & 747 		& $\quad G_\mu^\one$ & 696 \\
    $\quad\tilde{u}_L$ & 986 	& 
    \raisebox{-1.2ex}{$\quad Q_+^\one$} & \raisebox{-1.2ex}{662} \\[-.7ex]
    $\quad\tilde{d}_L$ & 989 	&& \\
    $\quad\tilde{u}_R$ & 942 	& $\quad U_-^\one$ & 608 \\
    $\quad\tilde{d}_R$ & 939 	& $\quad D_-^\one$ & 606 \\
    $\quad\neu_2$ & 224 		& $\quad Z_\mu^\one$ & 538 \\
    $\quad\neu_1$ & 119 		& $\quad B_\mu^\one$ & 487 \\
    $\quad\tilde{e}_L$ & 326 	& $\quad L_+^\one$ & 521 \\
    $\quad\tilde{e}_R$ & 164 	& $\quad E_-^\one$ & 508 \\
    $\quad\tilde{G}$ & \phantom{10}0& $\quad B_H^\one$ & 427 \\
    \hline
    \multicolumn{2}{|l|}{Branching ratios} & \multicolumn{2}{|l|}{Branching ratios} \\
    \hline
    BR[$\tilde{g} \to q\,\bar{q}\, \neu_2$] & \phantom{1}16\%
    &
    BR[$G_\mu^\one \to q\,Q_+^\one$] & 50\%
    \\
    && BR[$Q_+^\one \to q \,Z_\mu^\one$] & \phantom{0}6.4\% 
    \\
    BR[$\neu_2 \to e^+e^- \neu_1$] & \phantom{1}26\%
    &
    BR[$Z_\mu^\one \to e^+e^- B_\mu^\one$] & \phantom{0}1.5\%
    \\
    BR[$\neu_1 \to \gamma \, \tilde{G}$] & 100\%
    &
    BR[$B_\mu^\one \to \gamma \, B_H^\one$] & 34\%
    \\
    \hline
  \end{tabular}
\mycaption{Masses and branching fractions for the GMSB scenario G1a (left) and
the UED6 scenario U1 (right).}\label{tab:scen}
}

In universal extra dimensions, on the other hand, the masses of all particles of
one KK level $(j,k)$ have the same value $m_{\rm KK}^2 = (j^2+k^2)/R^2$, where
$R$ is the size of the extra dimensions. This degeneracy is lifted only by
radiative corrections \cite{6dsm,Ponton:2005kx}, which shift the masses by up to
20\%. As a result, a typical spectrum for the particles of KK level 1 in UED6 is
much less hierarchical than in GMSB. In our numerical analysis, we will use the
reference scenario U1, defined by $R^{-1} = 500$~GeV, with the masses and
branching ratios \cite{uedlhc} given in Tab.~\ref{tab:scen}.

These qualitative features of the spectra could be used to distinguish GMSB and
UED6 experimentally. However, there are several caveats to consider:
The mass spectra of GMSB can vary substantially
in non-minimal models, see {\it e.\hspace{.3ex}g.} Ref.~\cite{nmgmsb}.
Extra dimensional models are known to become strongly coupled at large energies
and require some new physics to be present at the scale. The effects of this
unknown high-scale physics could generate mass contributions to the KK particles
in UED \cite{Cheng:2002iz}. However, the spins of the the new particles can
serve as very robust discriminators between different models.

A non-zero spin of a particle can lead to angular correlations between its
decay products. At the LHC, angular correlations are manifested in the
invariant mass distributions of the visible decay products of a decay chain. 
The long decay chains \eqref{eq:longch1} and \eqref{eq:longch2} are
of the general form\begin{equation}
D \rightarrow l_n^\pm C \rightarrow l_n^\pm l_f^\mp B \rightarrow l_n^\pm l_f^\mp
\gamma A\,,
\label{eq:longchain}
\end{equation}
with $m_A < m_B < m_C < m_D$. Here we call the lepton that is emitted in
the first decay step the ``near-lepton'' $l_n$, while the lepton from the second
decay step is named the ``far-lepton'' $l_f$. From this one can construct the
invariant masses
\begin{align}
m_{n\gamma}^2 &\equiv (p_{l_n} + p_\gamma)^2 =
	\left(m_{n\gamma}^{\rm max}\right)^2 \, \frac{1}{4}
	\biggl [2-\Bigl (1-\frac{m_B^2}{m_C^2}	\Bigr )
		\left(1-\cos \theta_{nf}^{(C)}\right) \biggr]
		\left(1-\cos \theta_{n\gamma}^{(B)}\right),
\\
m_{f\gamma}^2 &\equiv (p_{l_f} + p_\gamma)^2 =	
	\left(m_{f\gamma}^{\rm max}\right)^2 \frac{1}{2} \left(1-\cos \theta_{f\gamma}^{(B)}\right),
\\
m_{nf}^2 &\equiv (p_{l_n} + p_{l_f})^2 =	
	\left(m_{nf}^{\rm max}\right)^2 \frac{1}{2} \left(1-\cos \theta_{nf}^{(C)}\right),
\\
m_{nf\gamma}^2 &\equiv (p_{l_n} + p_{l_f} + p_\gamma)^2 =
	m_{n\gamma}^2 + m_{f\gamma}^2 + m_{f\gamma}^2\,,
\end{align}
which are related to $\theta_{nf}^{(C)}$, the angle between the near-lepton and
the far-lepton in the rest frame of $C$, $\theta_{n\gamma}^{(B)}$,
the angle between the near-lepton and the photon in the $B$ rest frame,
and $\theta_{f\gamma}^{(B)}$, the angle between the far-lepton and the photon in
the $B$ rest frame, respectively.
The maximum values for the invariant masses are given in eq.~\eqref{eq:mmax} in
the appendix.

For a given decay matrix element, the distribution with respect to some
invariant mass is then obtained by integrating over all remaining kinematical
variables in a given reference frame, as described in detail in
Refs.~\cite{Miller:2005zp,ananda}.

In practice, the near and far leptons cannot be distinguished in a
straightforward way. Instead the observable lepton-photon invariant mass
distribution is the sum of $
{\rm d}\Gamma/{\rm d}m_{n\gamma}^2 + 
{\rm d}\Gamma/{\rm d}m_{f\gamma}^2$. 
Additional information can be obtained from the distributions with respect to
the minimum and maximum of the lepton-photon invariant masses,
\begin{align}
m_{h\gamma}^2 &= \max\{m_{n\gamma}^2,m_{f\gamma}^2\}, &
m_{l\gamma}^2 &= \min\{m_{n\gamma}^2,m_{f\gamma}^2\}.
\end{align}
Since the total magnitude of the decay width does not carry any information
about the spins of the particles involved, we will normalize the invariant mass
distributions to unity,
\begin{equation}
\frac{1}{\Gamma_0} \, \frac{{\rm d}\Gamma}{{\rm d}m} \equiv 
\frac{{\rm d}P}{{\rm d}m}\,,
\end{equation}
where $\Gamma_0$ is the integrated decay width of the given decay channel, and
${\rm d}P$ is defined as a differential probability density.

For the short decay chains \eqref{eq:shortch1} and \eqref{eq:shortch2}
of the general form
\begin{equation}
C \rightarrow q B \rightarrow q \gamma A\,,
\label{eq:shortchain}
\end{equation}
with $m_A < m_B < m_C$, the only observable invariant mass distribution that can
be constructed is
\begin{equation}
m_{q\gamma}^2 \equiv (p_q + p_\gamma)^2 =	
	\left(m_{q\gamma}^{\rm max}\right)^2 \frac{1}{2} \left(1-\cos \theta_{q\gamma}^{(B)}\right).
\end{equation}
Here $(m_{q\gamma}^{\rm max})^2 = (m_C^2-m_B^2)(m_B^2-m_A^2)/m_B^2$.

\subsection{Spin correlations in GMSB}

In GMSB the final state $l^+l^-\gamma+\ETmiss$ is fed by the decay chain
eq.~\eqref{eq:longch1}, with $\slR^\pm$ and $\neu_1$ as intermediate particles.
Since the $\slR^\pm$ is a scalar it does not transmit any angular correlations.
The fermionic $\neu_1$ can lead to non-trivial spin correlations in the decay
chain, but only if the couplings at both the production and decay vertex are
chiral, {\it i.$\,$e.} left- and right-handed components have different coupling
strength \cite{Wang:2006hk}. While this condition is fulfilled for the
$\slR^\pm$-$l^\mp$-$\neu_1$ vertex, the $\neu_1$-$\gamma$-$\tilde{G}$ vertex has
the form \cite{sugra}
\begin{equation}
\raisebox{-1.3cm}{\epsfig{figure=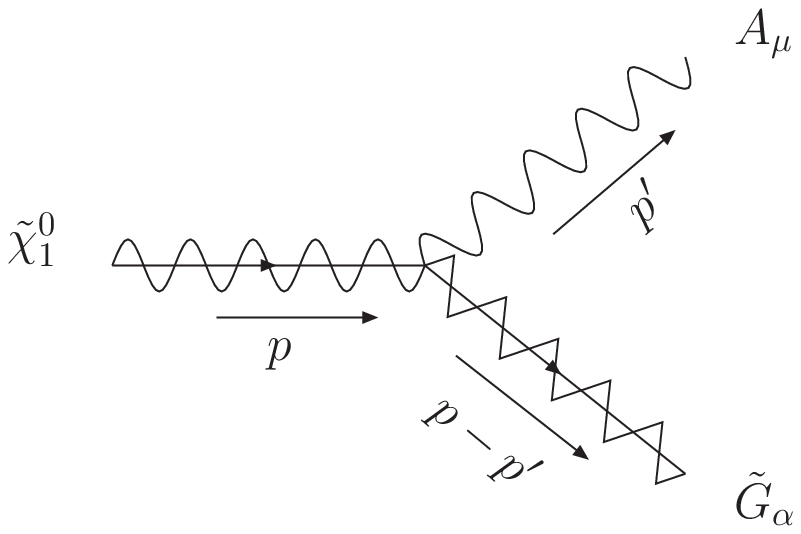, width=4.5cm}}
\propto \left[\pslash',\gamma_\mu \right] \gamma_\alpha.
\end{equation}
Here the left- and right-handed component of the neutralino couple with equal
strength so that no observable angular correlations are generated.
Thus the only features in the invariant mass distributions of the GMSB 
decay chain \eqref{eq:longch1} are generated by the phase space.
Analytical results for the lepton-photon and lepton-lepton distributions 
are listed in the appendix.

Similarly, the short decay chain eq.~\eqref{eq:shortch1} does not lead to
visible angular correlation effects. The quark-photon invariant mass
distribution is given by
\begin{equation}
\frac{{\rm d}P}{{\rm d}m_{q\gamma}^2} =
\frac{m_B^2}{(m_C^2-m_B^2)(m_B^2-m_A^2)}\,.
\label{eq:shortgmsb}
\end{equation}
By comparison with the formulas in the appendix one can see that
eq.~\eqref{eq:shortgmsb} is identical to the $m_{f\gamma}^2$ distribution of the
long chain. This can be easily understood by the fact that the chirality of the
slepton and squark couplings is identical.

\subsection{Spin correlations in UED6}

The typical mass hierarchy generated by radiative corrections in UED6,
$m_{Z_\mu^\one} > m_{L_+^\one} > m_{B_\mu^\one} > m_{B_H^\one}$, enables the
decay chain eq.~\eqref{eq:longch2}. However, in general sizeable corrections to
the KK-particle masses could be generated by the unknown physics that complete
the theory at high energies \cite{Cheng:2002iz}. Thus for completeness we will
study all possible decay chains that, for suitable mass hierarchies, could lead
to the final state $l^+l^-\gamma+\ETmiss$, as illustrated in
Fig.~\ref{fig:uedchains},
\FIGURE[ht]{
  \centering
  \begin{tabular}{cc} 
    \epsfig{figure=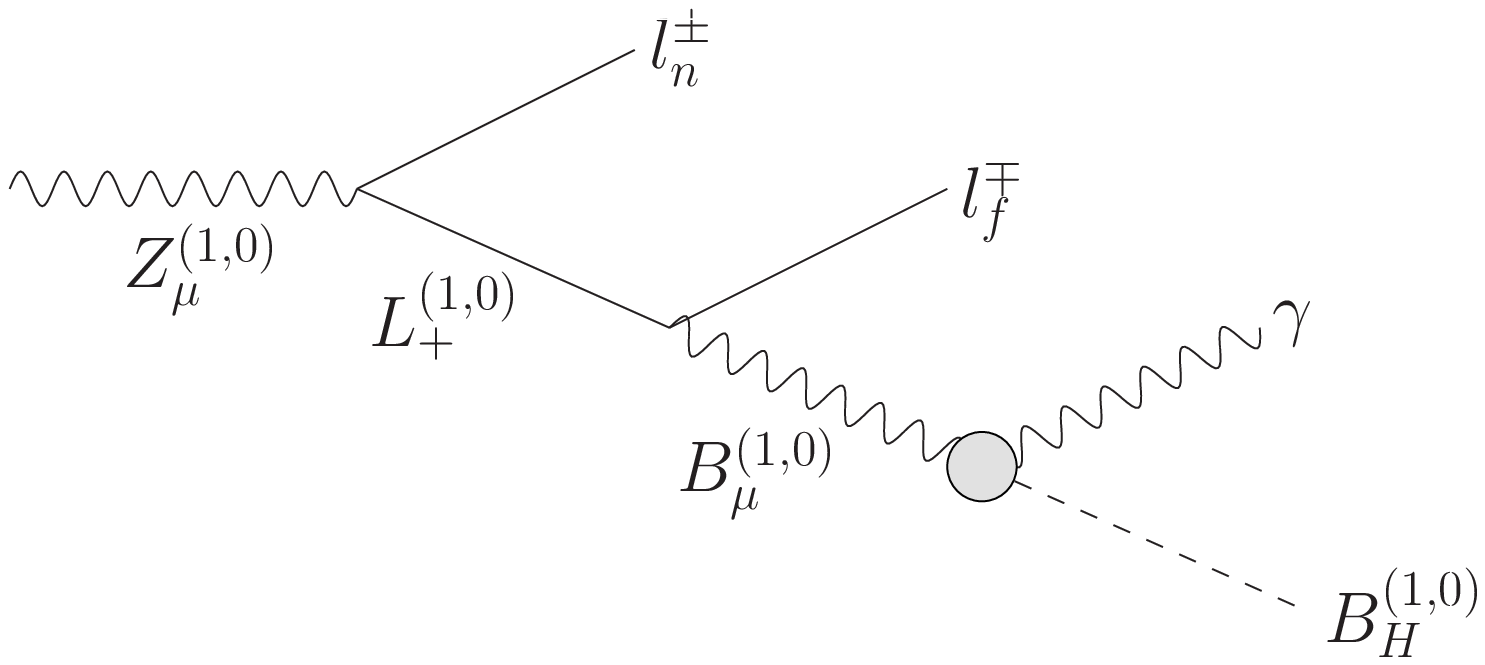, width=0.48\textwidth} &
    \epsfig{figure=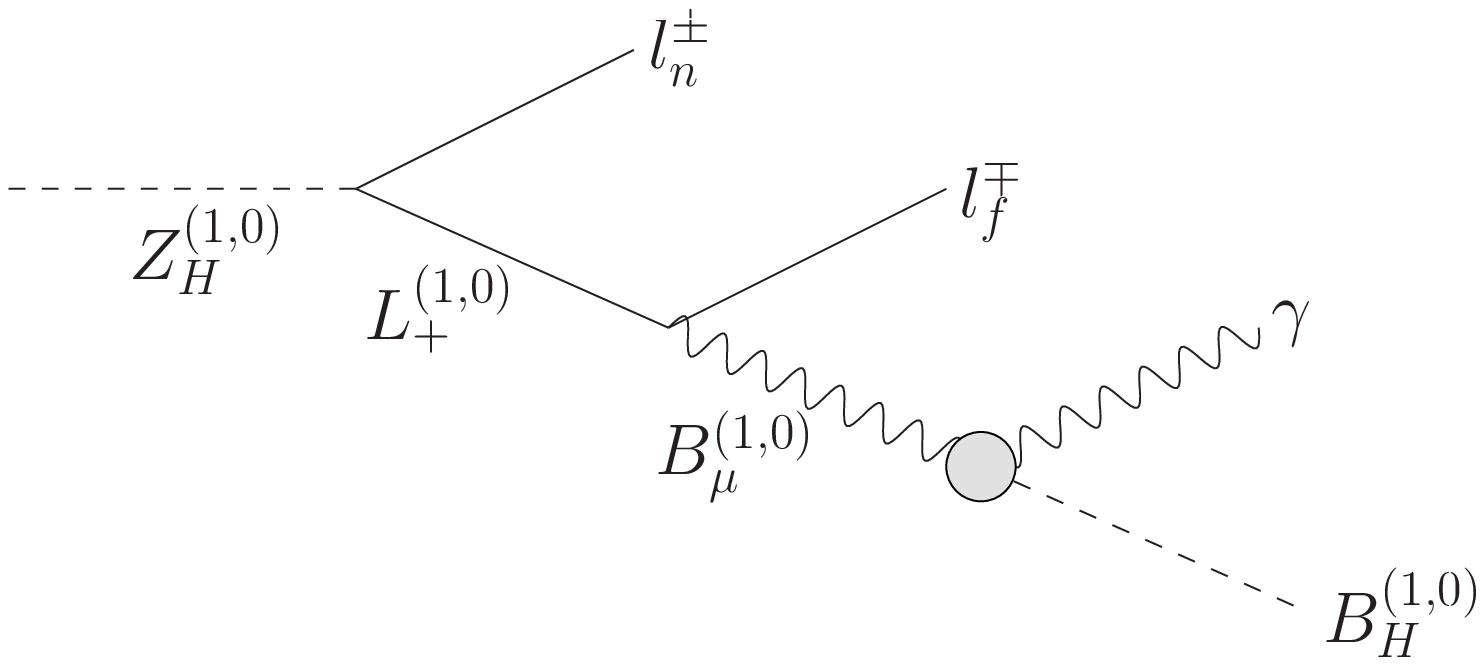, width=0.48\textwidth} \\[-1ex] 
    (VFVS) & (SFVS) \\[1em] 
    \epsfig{figure=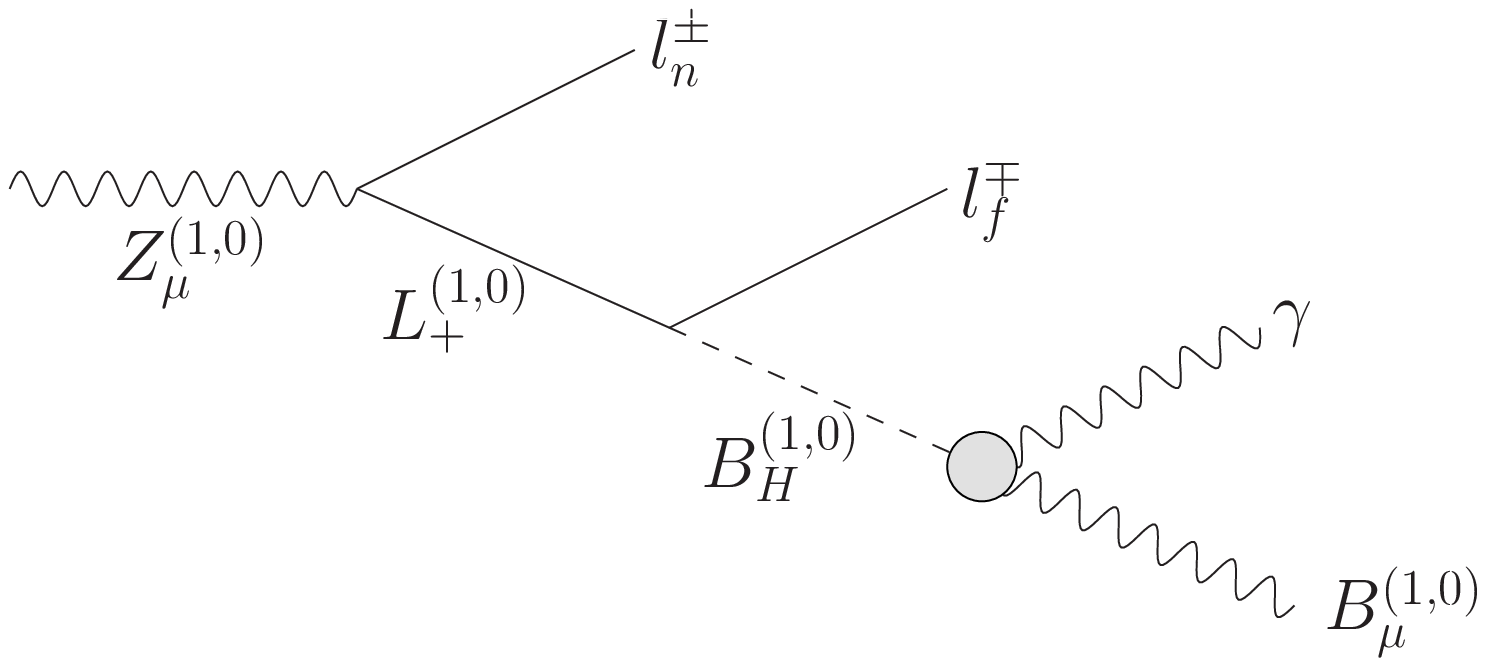, width=0.48\textwidth} &
    \epsfig{figure=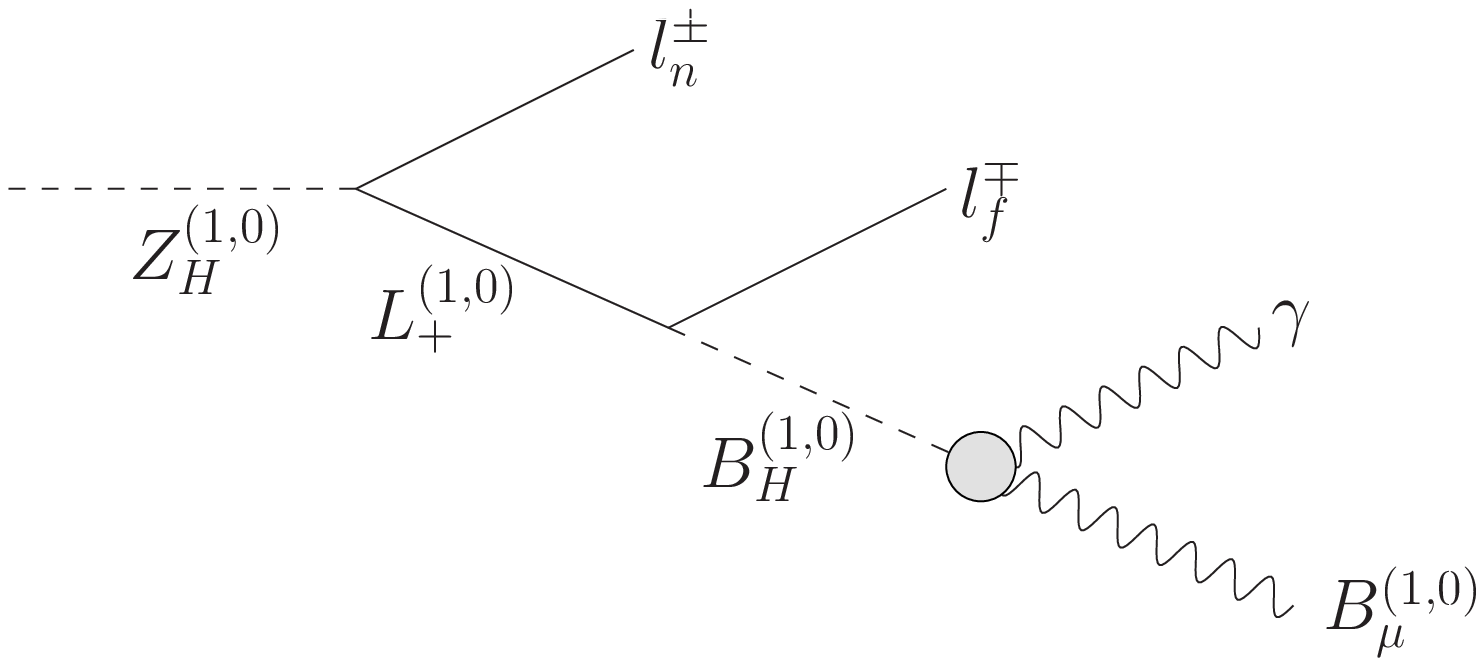, width=0.48\textwidth} \\[-1ex] 
    (VFSV) & (SFSV)  
  \end{tabular}
  \mycaption{Different UED6 decay chains with various spin configurations
    leading to the same
    $l^+l^-\gamma+\ETmiss$ signature.}\label{fig:uedchains}
}
\begin{align}
&Z_\mu^\one \to l^\pm \, L_+^\one \to l^+ l^- \, B_\mu^\one \to
l^+ l^- \, \gamma \, B_H^\one && {\rm (VFVS)}, \label{eq:ued1} \\
&Z_H^\one \to l^\pm \, L_+^\one \to l^+ l^- \, B_\mu^\one \to
l^+ l^- \, \gamma \, B_H^\one && {\rm (SFVS)}, \label{eq:ued2} \\
&Z_\mu^\one \to l^\pm \, L_+p^\one \to l^+ l^- \, B_H^\one \to
l^+ l^- \, \gamma \, B_\mu^\one && {\rm (VFSV)}, \label{eq:ued3} \\
&Z_H^\one \to l^\pm \, L_+^\one \to l^+ l^- \, B_H^\one \to
l^+ l^- \, \gamma \, B_\mu^\one && {\rm (SFSV)}. \label{eq:ued4}
\end{align}
Here we have introduced short-hand notations for the four decay chains based on
the KK particles at each decay stage being a 
scalar (S), fermion (F) or vector (V).
In all cases we keep the couplings of the KK particles as they are predicted by
the UED6 model, and mixing between gauge eigenstates is neglected.
Analytical results for the invariant mass distributions for all four
combinations are listed in the appendix.

In case of the short decay chain, eq.~\eqref{eq:shortch2}, there are two
possible decay chains with the same final state, depending on the
mass hierarchy,
\begin{align}
&Q_-^\one \to q \, B_\mu^\one \to q \, \gamma \, B_H^\one && {\rm (FVS)}, \\
&Q_-^\one \to q \, B_H^\one \to q \, \gamma \, B_\mu^\one && {\rm (FSV)}.
\end{align}
The quark-photon invariant mass distributions for the two cases read
\begin{align}
{\rm FVS:} &\nonumber \\
  \frac{{\rm d}P}{{\rm d}m_{q\gamma}^2} & = 
\frac{3 m_B^4 \left(2 m_{q\gamma}^4 m_B^2-2 m_{q\gamma}^2 (m_A^2-m_B^2) (m_B^2-m_C^2)+(m_A^2-m_B^2)^2
(m_B^2-m_C^2)\right)}{(m_A^2-m_B^2)^3 (m_B^2-m_C^2)^2 (2 m_B^2+m_C^2)},\\
{\rm FSV:} & \nonumber \\
  \frac{{\rm d}P}{{\rm d}m_{q\gamma}^2} & = 
	\frac{m_B^2}{(m_C^2-m_B^2)(m_B^2-m_A^2)}\,.
\end{align}
As before, one can see that these are identical to the $m_{f\gamma}^2$
distributions of the long chain for the VFVS/SFVS and VFSV/SFSV combinations,
respectively. In the FSV case, since the intermediate $B$ particle is a scalar,
this decay process does not involve any spin correlations and is identical to
the pure phase space distribution, and thus to the GMSB distribution.

\subsection{Discussion of analytical results}

\FIGURE[ht]{
  \centering
  \begin{tabular}{cc} 
    \epsfig{figure=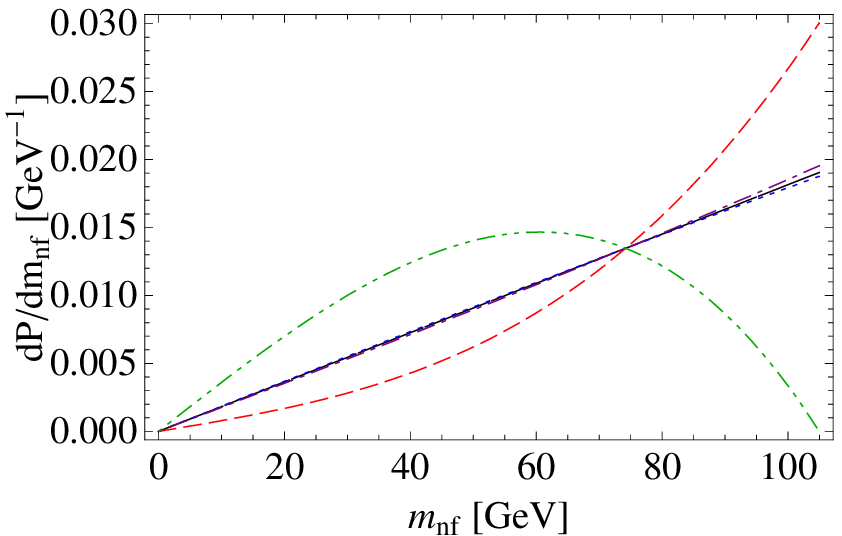, width=0.48\textwidth} &
    \epsfig{figure=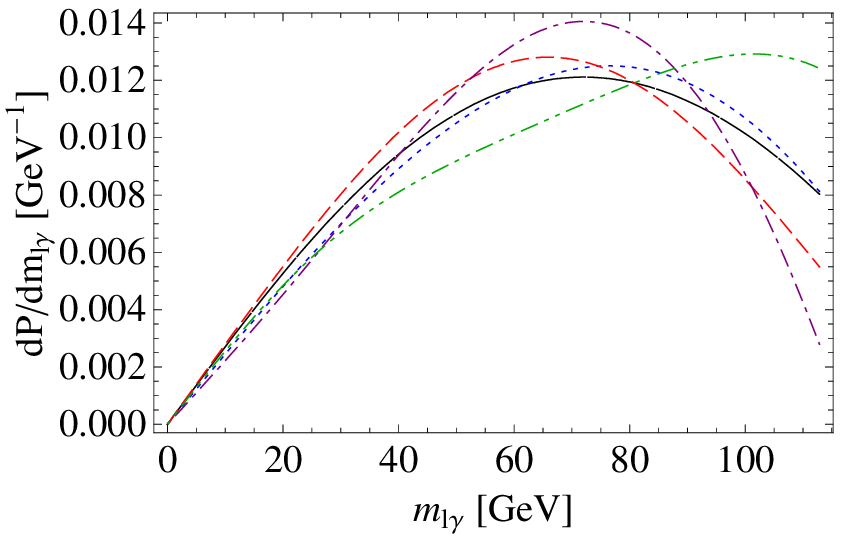, width=0.48\textwidth} \\
    \epsfig{figure=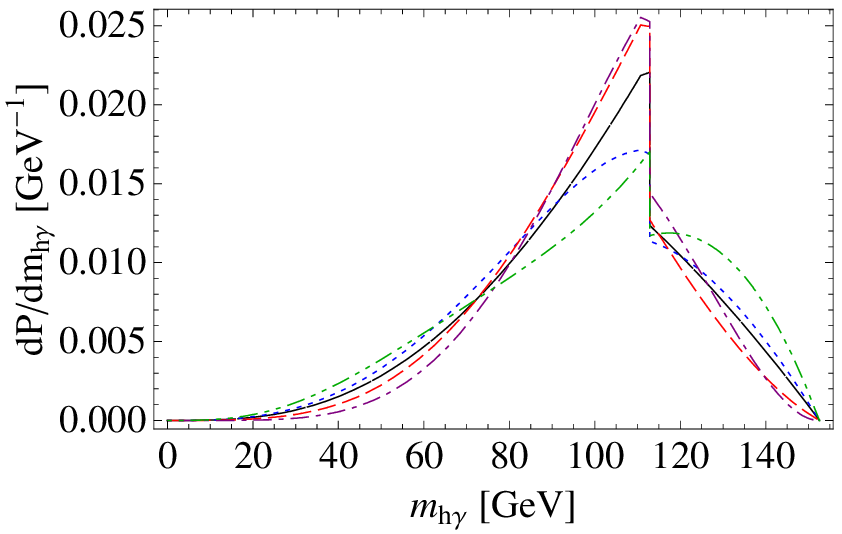, width=0.48\textwidth} &
    \epsfig{figure=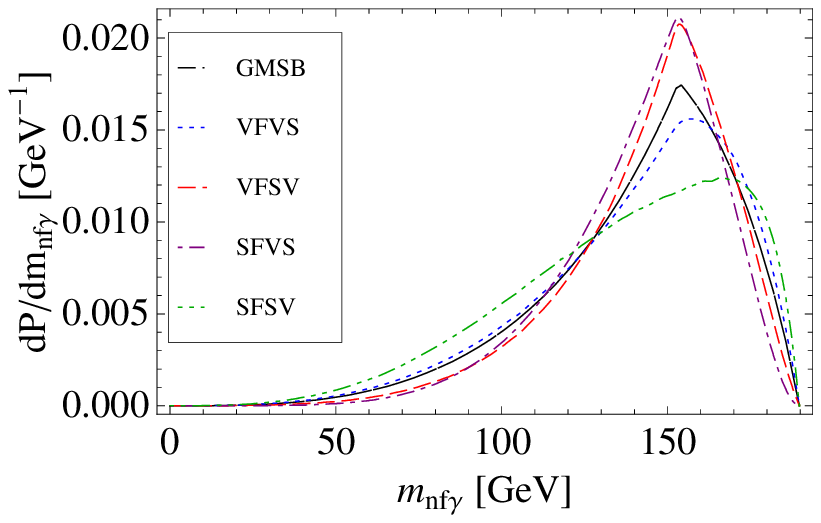, width=0.48\textwidth} \\
  \end{tabular}
  \mycaption{Observable invariant mass distributions for the decay chain in
    eq.~\eqref{eq:longchain} for different models and masses from the G1a scenario.}
  \label{fig:dmsusy}
}
\FIGURE[ht]{
  \centering
  \begin{tabular}{cc} 
    \epsfig{figure=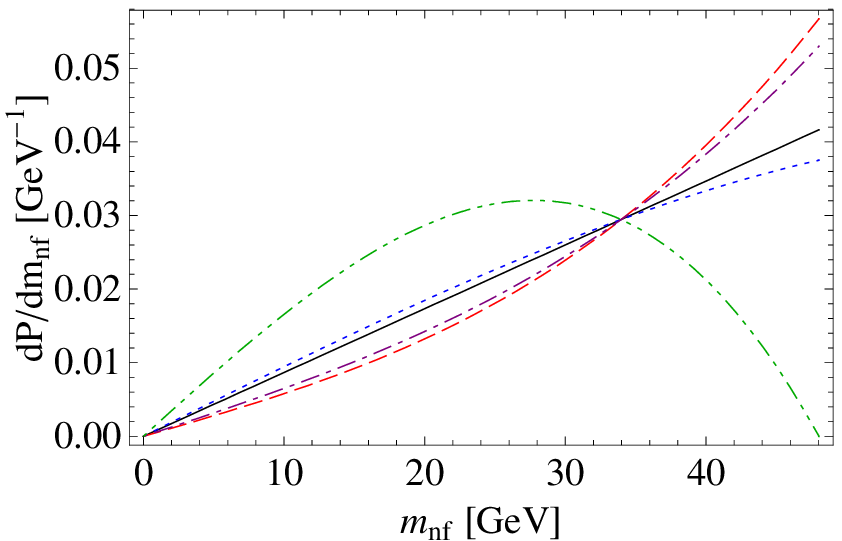,width=0.48\textwidth} &
    \epsfig{figure=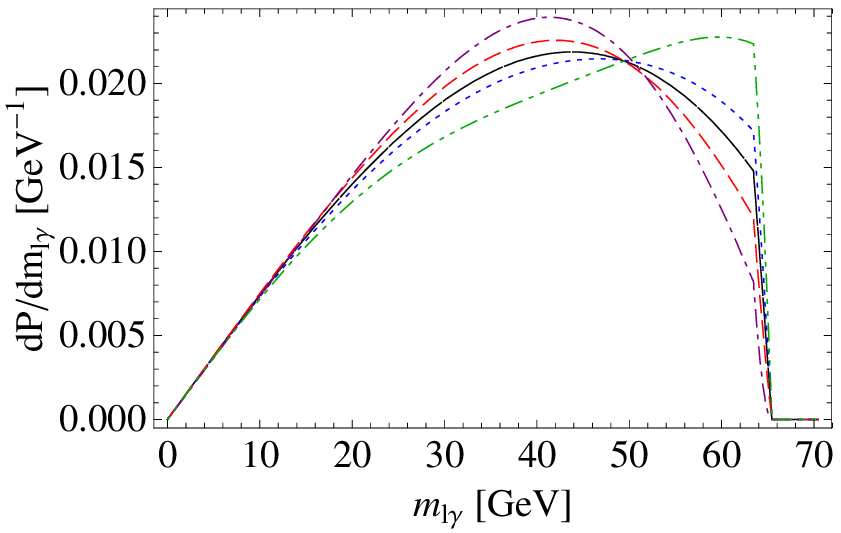,width=0.48\textwidth} \\
    \epsfig{figure=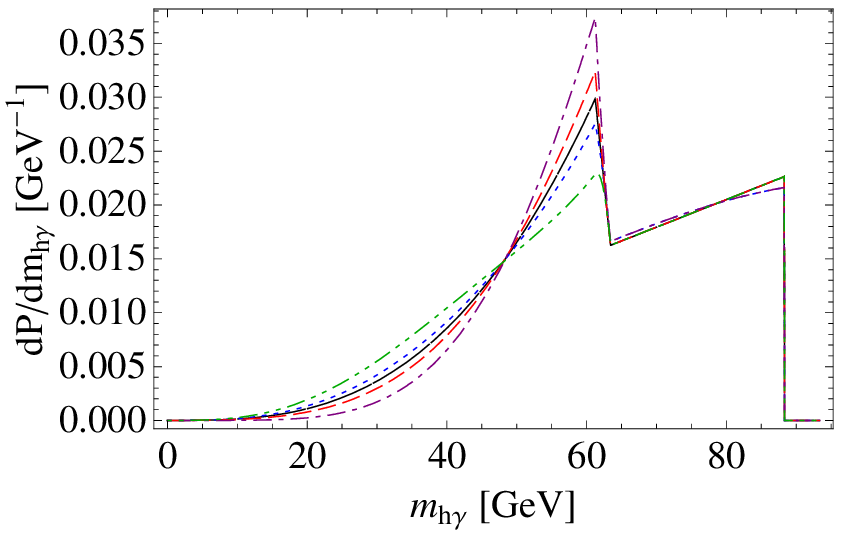,width=0.48\textwidth} &
    \epsfig{figure=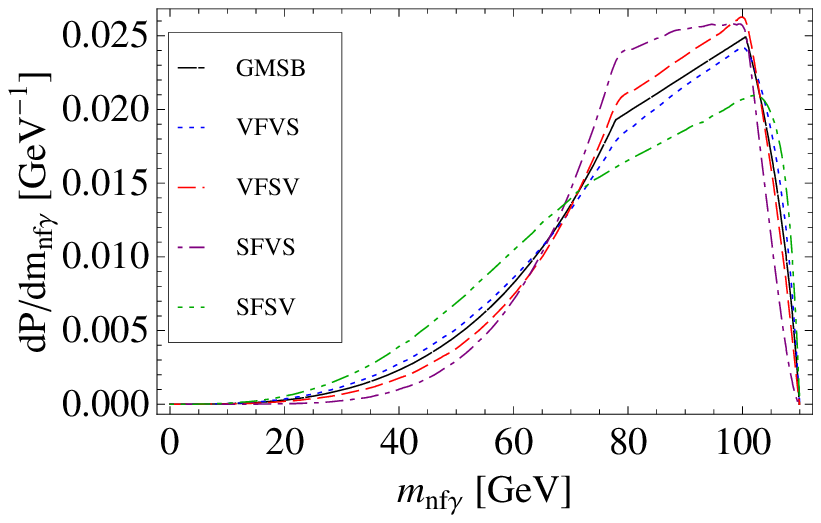,width=0.48\textwidth} \\
  \end{tabular}
  \mycaption{Observable invariant mass distributions for the decay chain in
    eq.~\eqref{eq:longchain} for different models and masses from the
    U1 scenario.}
  \label{fig:dmued}
}

\FIGURE[hb]{
  \centering
  \begin{tabular}{cc} 
    \epsfig{figure=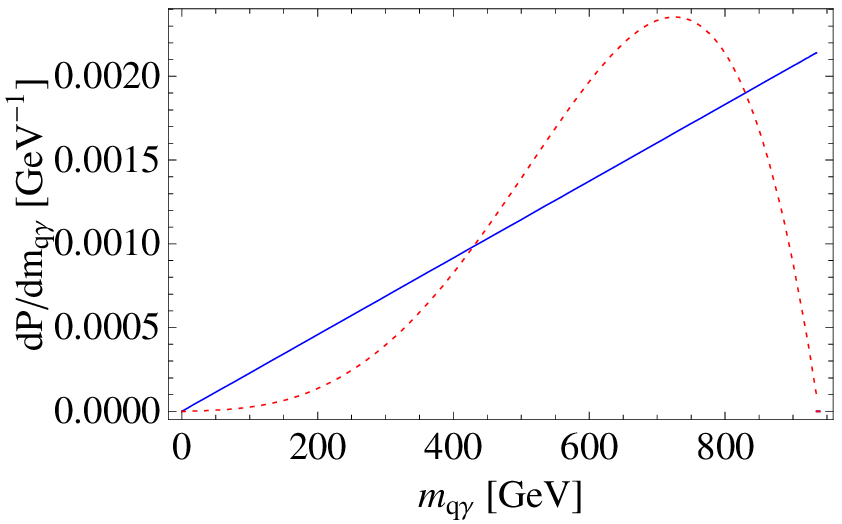,width=0.48\textwidth} &
    \epsfig{figure=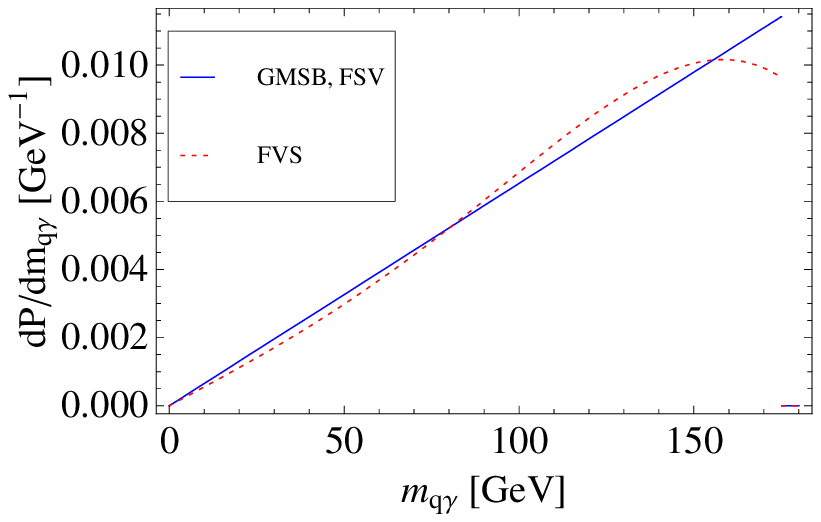,width=0.48\textwidth} \\
  \end{tabular}
  \mycaption{Observable invariant mass distributions for the short decay chain in
    eq.~\eqref{eq:shortchain} for masses from the G1a scenario (left)
    and U1 scenario (right).}\label{fig:dshort}
}

Figs.~\ref{fig:dmsusy} and \ref{fig:dmued} show the distributions for the four
independent observable invariant mass combinations of the $l^+l^-\gamma+\ETmiss$
final state: the di-lepton invariant mass $m_{nf}$, the ``low'' and ``high''
lepton-invariant masses $m_{l\gamma}$ and $m_{h\gamma}$, respectively, and the
lepton-lepton-photon invariant mass $m_{nf\gamma}$.  Each plot contains five
curves corresponding to the five models (or spin assignments) GMSB, VFVS, SFVS,
VFSV, and SFSV. In case of Fig.~\ref{fig:dmsusy}, for all five models
the masses have been chosen from the G1a scenario, with $m_A = m_{\tilde{G}} =
0$, $m_B=\mneu{1}$,
$m_C=m_{\tilde{e}_R}$, and $m_D=\mneu{2}$.
On the other hand, Fig.~\ref{fig:dmued} shows the situation for the U1 spectrum
with $m_A=m_{B_H^{(1,0)}}$,
$m_B=m_{B_\mu^{(1,0)}}$, $m_C=m_{L_+}^{(1,0)}$, and $m_D=m_{Z_\mu^{(1,0)}}$.

As evident from the plots, different distributions could discriminate between
different spin assignments. The di-lepton distribution ${\rm d}P/{\rm d}m_{nf}$
is markedly different for the VFSV and SFSV models, where the $B$ particle is a
scalar, compared to the other models. This can be understood from the fact that the
chiral structure of the KK-fermion couplings  to KK-scalars or KK-vector
bosons is different (see for example Ref.~\cite{kwy} for more details). On the
other hand, the peak in the ${\rm d}P/{\rm d}m_{h\gamma}$ and ${\rm d}P/{\rm
d}m_{nf\gamma}$ spectra is relatively enhanced for the VFSV and SFVS models.
As a result, the discriminative power between different spin assignments is
maximized by including all four distributions in the analysis.

Fig.~\ref{fig:dshort} shows the jet-photon invariant mass distribution for the
short decay chain. For the distribution on the left hand side the masses have been
chosen from the G1a scenario, $m_A = m_{\tilde{G}}=0$,
$m_B=m_{\tilde{\chi}_1^0}$, $m_C=m_{\tilde{u}_R}$ and on the right hand side the
distribution is shown for the U1 scenario, where $m_A = m_{B_H^{(1,0)}}$,
$m_B=m_{B_\mu^{(1,0)}}$, $m_C=m_{Q_-^{(1,0)}}$. As mentioned before, the GMSB
and FSV chains both do not generate any spin correlations and thus cannot be
distinguished from each other. However,
the plots show a discrimination potential between FVS and the other two cases,
especially for the G1a mass scenario.
The qualitative features of the jet-photon distribution are similar for the G1a and U1
scenarios, although the spin correlation effects are less pronounced for the
more degenerate U1 mass spectrum than for the more hierarchical G1a spectrum.


\section{Monte-Carlo simulation and numerical analysis}
\label{sc:mc}

In the previous section theoretical formulas for the invariant mass spectra of
decay chains in GMSB and UED6 were discussed. However, in a realistic experimental
setup one must take into account several effects that influence the measured
distributions. 
\begin{itemize}
\item
Detector effects, such as limited resolution, and
reconstruction effects from particle identification 
can smear out the visible invariant mass distributions.
Further, detector acceptance can distort the shape of the distribution.
\item
For the separation of the new physics signal from the SM
backgrounds suitable selection cuts have to be implemented, which can distort
the shape of the distributions.
\item
Since both SUSY and KK particles are produced in pairs, often there will be two
hard photons in one event, leading to a two-fold ambiguity in the reconstruction
of the correct decay chain.
\end{itemize}
Therefore we have performed a realistic experimental simulation of the new
physics signal from the long decay chain for GMSB and UED6, with the goal of
reconstructing the spin-sensitive invariant mass distributions and
discriminating between models  from the simulated data.

The decay chains in eqs.~\eqref{eq:longch1}, \eqref{eq:ued1}--\eqref{eq:ued4}
have been computed with \textsc{CompHEP 4.4} \cite{comphep} for 14~TeV
proton-proton collisions, using the GMSB model
file from Ref.~\cite{gmsb} and our own implementation of the UED6
model. Parton-level events generated with \textsc{CompHEP} were then passed on
to \textsc{Pythia 6.4.12} \cite{pythia}. 
Finally, the ATLAS detector\cite{atlas} is simulated using the
parametrized fast detector simulation ATLFAST\cite{atlfast1,atlfast2},
which includes detector acceptance, resolution and some basic particle identification.

For the G1a and U1 mass spectra, the aforementioned decay chains are initiated
mostly by squark (KK-quark) and gluino (KK-gluon) production processes,
respectively. Since our analysis does not depend crucially on the details of
the hadronic decay products of the squarks (KK-quarks) and gluinos (KK-gluons)
we have only generated events for squark (KK-quark) pair production as the
primary hard process, and then normalized the total event count according to the
total production cross section including gluino (KK-gluon) processes.

As a further simplification, we have only generated matrix elements with the
decay chain of one of the two squarks (KK-quarks) within \textsc{CompHEP}, which
correctly implements all spin correlations. The decay of the second squark
(KK-quark) was simulated in \textsc{Pythia}, without spin correlations.
Nevertheless, this procedure provides a good approximation to the complete
matrix elements since the branching ratio for the decay chain leading to the
final state $l^+l^-\gamma$  is relatively small and thus only a very small
fraction of the events contains two decay cascades of this type. Moreover,
\textsc{Pythia} was used for simulating initial and final state radiation and
hadronization.

The event selection was performed according to Ref.~\cite{hp}. First the
effective mass
\begin{equation}
M_{\text{eff}} = \ETmiss + p_{\top,1} + p_{\top,2} + p_{\top,3} + p_{\top,4}
\end{equation}
is defined, where $\ETmiss$ denotes the missing transverse energy and $p_{\top,i}$ the
transverse momenta of the 4 hardest jets. The selected events need to
fulfill the following conditions:
\begin{enumerate}
\item 4 jets with transverse momenta $> 25 \; \text{GeV}$,
\item $M_{\text{eff}} > 400 \; \text{GeV}$,
\item $\ETmiss > 0.1 \, M_{\text{eff}}$,
\item 2 photons with transverse momenta $> 20 \; \text{GeV}$,
\item 2 electrons or muons with transverse momenta $> 20 \; \text{GeV}$.
\end{enumerate}
After application of these cuts the SM background is reduced to a negligible
level \cite{hp}, while about 20\% of the signal is retained.

Since the selected events include two photons, one of them has to be selected to
compute the invariant mass distributions. Good results are obtained, when choosing the one photon, which gives the smaller $m_{nf\gamma}^2$ invariant mass.

It would be interesting to also analyze the short decay chain in a complete
simulation, since the expected rates are large (about 5 pb both for G1a and
U1). However, the signature of this final state, two hard jets, two
photons, and missing energy, is very sensitive to issues related to jets faking
photons. Therefore, it would require a more careful analysis of QCD backgrounds, 
which will be left for a future publication.

\subsection{Cross sections}

For a realistic analysis, the model-dependent cross-sections and event
numbers have to be calculated. At the G1a point, the total squark and gluino
production cross section is $\sigma_{\tilde{q}/\tilde{g}}=7.6 \; \text{pb}$
\cite{hp}. Since the gluino is lighter than the squarks, the squarks decay through
cascades involving a gluino. With the branching ratios from Tab.~\ref{tab:scen} 
a good approximation for the cross section for the decay chain is
\begin{align}
\sigma_{\text{G1a}} &= \sigma_{\tilde{q}/\tilde{g}} \times 2 \times 2 \times
\text{BR}\left[ \tilde{g} \rightarrow q \bar{q} \tilde{\chi}_2^0 \right] \times
\text{BR} \left[ \tilde{\chi}_2^0 \rightarrow e^+ e^- \tilde{\chi}_1^0 \right]
\times \text{BR} \left[ \tilde{\chi}_1^0 \rightarrow \gamma \tilde{G} \right]
\nonumber \\ &\simeq 1.2 \; \text{pb}.
\end{align}
The factors of $2$ result from the two generations of leptons in the di-leptonic decay
chain and the fact that squarks or gluinos are produced in pairs and both of them
can decay through this channel.

In the U1 model, the KK-quarks and KK-gluons are produced with the following
cross sections \cite{uedlhc}
\begin{align}
&\sigma_{Q_+^{(1,0)} Q_+^{(1,0)}} \sim 7 \; \mathrm{pb}, & &\sigma_{Q_+^{(1,0)} Q_-^{(1,0)}} \sim 18 \; \mathrm{pb}, \nonumber \\
&\sigma_{G_\mu^{(1,0)} G_\mu^{(1,0)}} \sim 10 \; \mathrm{pb}, & &\sigma_{G_\mu^{(1,0)} Q_+^{(1,0)}} \sim 24 \; \mathrm{pb}, \nonumber \\
&\sigma_{G_\mu^{(1,0)} Q_-^{(1,0)}} \sim 26 \; \mathrm{pb}.
\end{align}
In this scenario, KK-quarks are lighter than the KK-gluon, such that the latter will
mostly decay into KK-quarks. Therewith, the total KK-quark production cross section is
\begin{align}
\sigma_{Q_+^{(1,0)}}&=2 \: \sigma_{Q_+^{(1,0)}Q_+^{(1,0)}} + \sigma_{Q_+^{(1,0)}Q_-^{(1,0)}} + 2 \: \mathrm{BR}\left[G_\mu^{(1,0)} \rightarrow Q_+^{(1,0)} q \right] \sigma_{G_\mu^{(1,0)} G_\mu^{(1,0)}}   \\
&\quad+ \left( 1 + \mathrm{BR}\left[G_\mu^{(1,0)} \rightarrow Q_+^{(1,0)} q
\right] \right)\: \sigma_{G_\mu^{(1,0)} Q_+^{(1,0)}} +
\mathrm{BR}\left[G_\mu^{(1,0)} \rightarrow Q_+^{(1,0)} q \right] \:
\sigma_{G_\mu^{(1,0)} Q_-^{(1,0)}}.
\nonumber
\end{align}
As above, factors of 2 account for the two sides of the pair production process.
Then the total cross section for the decay chain adds up to
\begin{align}
\sigma_{\text{U1}} &= \sigma_{Q_+^{(1,0)}} \times 2 \times 2 \times \text{BR}\left[Q_+^{(1,0)} \rightarrow q Z_\mu^{(1,0)} \right] \times \text{BR}\left[ Z_\mu^{(1,0)} \rightarrow e^+ e^- B_\mu^{(1,0)} \right] \nonumber \\
&\quad \times \text{BR} \left[ B_\mu^{(1,0)} \rightarrow \gamma B_H^{(1,0)} \right] \simeq 0.12 \; \text{pb}.
\end{align}
Using the signal efficiency of the cuts in Ref.~\cite{hp} and assuming an
integrated luminosity of 10 fb$^{-1}$
we obtain $\text{N}_{\text{G1a}}^{(10)}=2500$ selected events for the G1a point and
$\text{N}_{\text{U1}}^{(10)}=250$ selected events for the U1 model.

It should be noted that the event rates depend strongly on the underlying model
and its parameters, as well as the choice of selection cuts. Furthermore, the
signal efficiency after cuts depends also on the decay chain of the second
squark or KK-quark, whose branching ratios vary between different models. For
the purpose of this study, we do not vary the choice of cuts, cross sections and
signal efficiency between different models when we compare them for one given
mass spectrum (i.~e.\ for all spin assignments we assume 2500 selected events for the G1a
mass spectrum and 250 selected events for the U1 mass spectrum). Rather, our numerical
analysis of the two scenarios should only serve as concrete examples for a spin
determination of a new physics signal, in particular since our method does not rely
on information about total rates.

\subsection{$\chi^2$ analysis}

In order to discriminate the histograms for the different spin configurations we
used the $\chi^2$-test implemented in ROOT \cite{root}. It
returns the $\chi^2$-probability, i.~e. the probability that two histograms with
identical underlying  distribution functions have a bigger $\chi^2$ value than
the two compared ones. Since these values depend on the number of bins, the discrimination was performed with 5 bins, which showed the best discriminative power at $10 \; \text{fb}^{-1}$. In Tab.~\ref{tab:ued} the minimal $\chi^2$ probabilities for each
pair of spin configurations are listed.

\TABLE[ht]{
  \centering
  \begin{tabular}{|r|lllll|}
    \hline
    & GMSB & VFVS & VFSV & SFVS & SFSV \\
    \hline
    GMSB &  & \cellcolor{lblue} 0.000 ($m_{h\gamma}$) & \cellcolor{lblue} 0.000 ($m_{nf}$) & \cellcolor{lblue} 0.006  ($m_{h\gamma}$) & \cellcolor{lblue} 0.000 ($m_{nf}$) \\
    
    VFVS & \cellcolor{lred} 0.056 ($m_{h\gamma}$) & & \cellcolor{lblue} 0.000 ($m_{nf}$) & \cellcolor{lblue} 0.000 ($m_{h\gamma}$) & \cellcolor{lblue} 0.000 ($m_{nf}$) \\
    
    VFSV & \cellcolor{lred} 0.577 ($m_{nf}$) & \cellcolor{lred} 0.155 ($m_{h\gamma}$) & & \cellcolor{lblue} 0.000 ($m_{nf}$) & \cellcolor{lblue} 0.000 ($m_{nf}$) \\
    
    SFVS & \cellcolor{lred} 0.025 ($m_{l\gamma}$) & \cellcolor{lred} 0.065 ($m_{l\gamma}$) & \cellcolor{lred} 0.084 ($m_{h\gamma}$) & & \cellcolor{lblue} 0.000 ($m_{nf}$) \\
    
    SFSV & \cellcolor{lred} 0.000 ($m_{nf}$) & \cellcolor{lred} 0.000 ($m_{nf}$) & \cellcolor{lred} 0.000 ($m_{nf}$) & \cellcolor{lred} 0.000 ($m_{nf}$) & \\
    \hline
  \end{tabular}
  \mycaption{Minimal $\chi^2$-probabilities for $10 \; \text{fb}^{-1}$. The
    distributions that provide the strongest constraints are noted in parentheses.
    The values in the blue cells (upper right) are for the G1a mass spectrum and in
    the red cells (lower left)  are the values for the U1 mass spectrum.}
  \label{tab:ued}
}

The results of the $\chi^2$ probabilities reflect the general features that can be seen in the histograms of
Fig.~\ref{fig:histsusy} and \ref{fig:histued}. As expected, the
discrimination between different spin combinations is far more effective for the
G1a scenario than for the U1 scenario, as a result of the more degenerate mass
spectrum and lower cross section in the latter case. For the G1a scenario,
even with 10~fb$^{-1}$ almost all models can be distinguished with a confidence
level better than 99.9\%. In the case of the U1 scenario only the SFSV spin
assignment can be distinguished at this confidence level with 10~fb$^{-1}$ luminosity,
owing to the distinctly different shape of the distribution of the
di-lepton invariant mass $m_{nf}$. In other cases however, in particular for the
GMSB spin assignment and the extra-dimensional chain VFSV,
it is not possible to make a distinction even at the 95\%
confidence level. Here a much larger integrated luminosity would be required
for a significant discrimination. 

As an example Fig.~\ref{fig:histsusy30} shows the reconstructed
invariant mass distributions corresponding to G1a masses for an
integrated luminosity of 30~fb$^{-1}$. The distributions are divided in
10 bins, which gives the best discriminative power for this
luminosity. More shape information is available due to more bins and
events allowing for a better separation between the different models.
For the U1 mass scenario the GMSB spin assignment can be distinguished from the
extra-dimensional chain VFSV at almost 90\% confidence level.

\FIGURE[ht]{
  \centering
  \begin{tabular}{cc} 
    \epsfig{figure=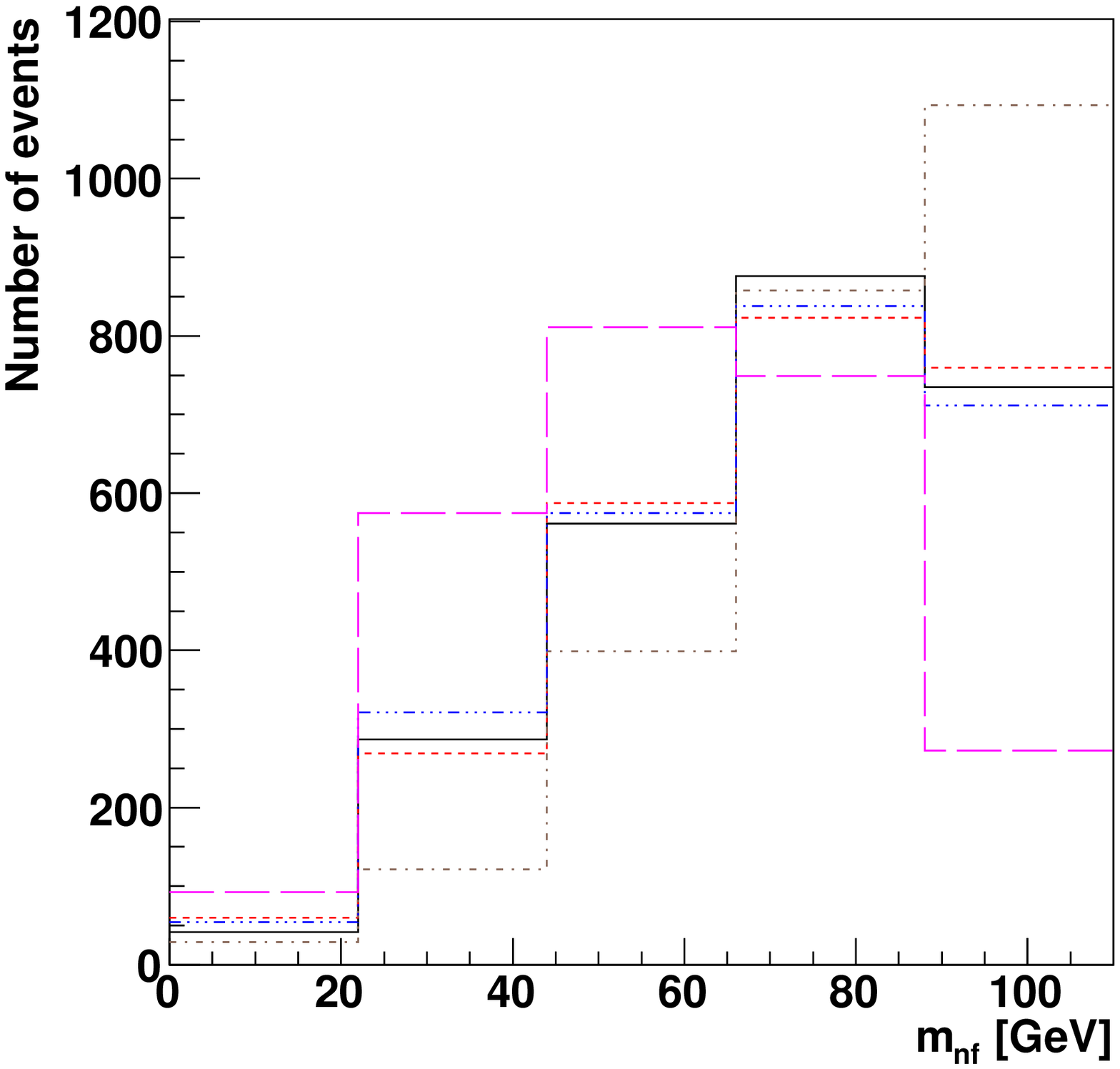,width=0.48\textwidth, bb=0 0 567 480, clip=true} &
    \epsfig{figure=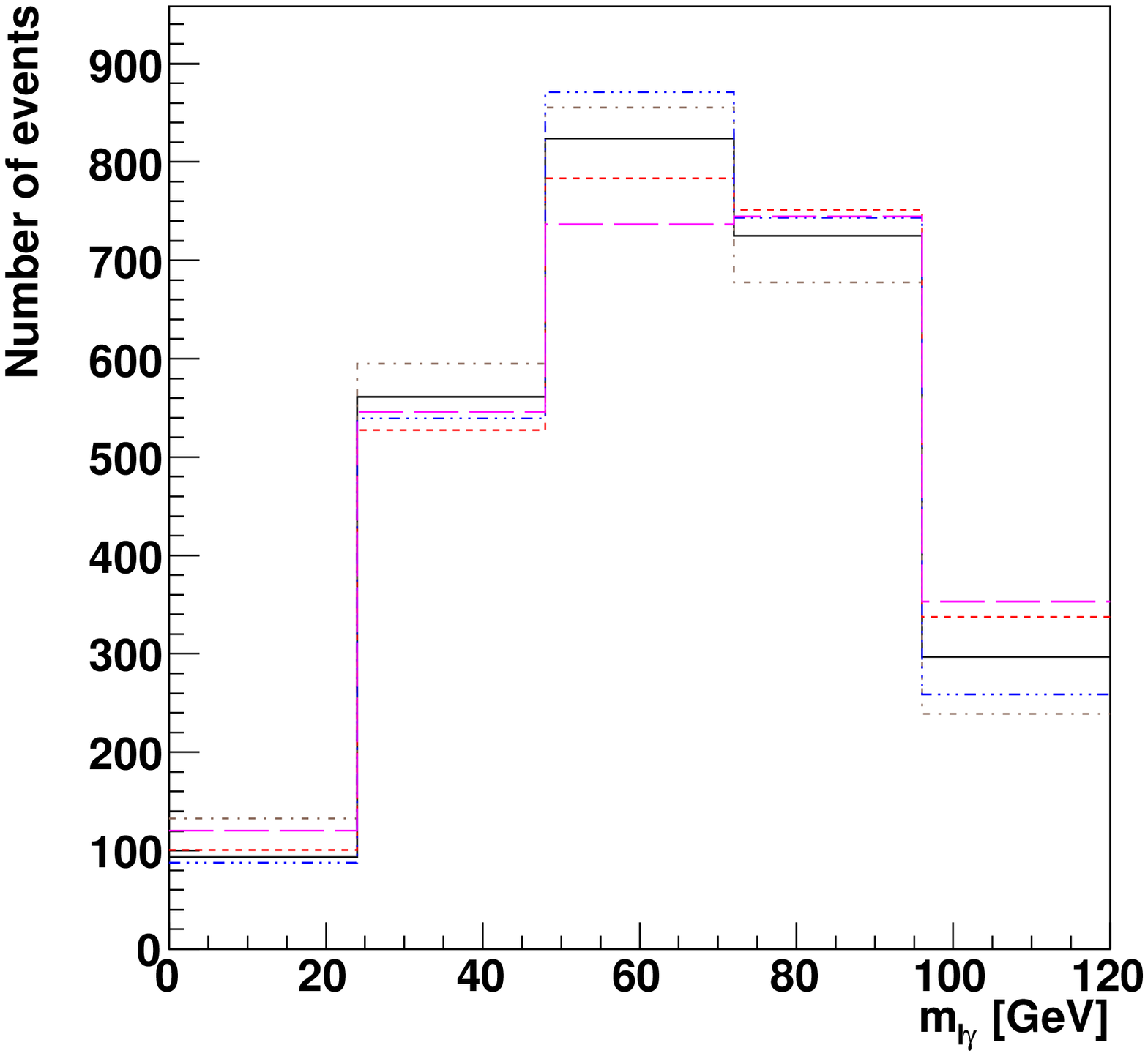,width=0.48\textwidth, bb=0 0 567 480, clip=true} \\
    \epsfig{figure=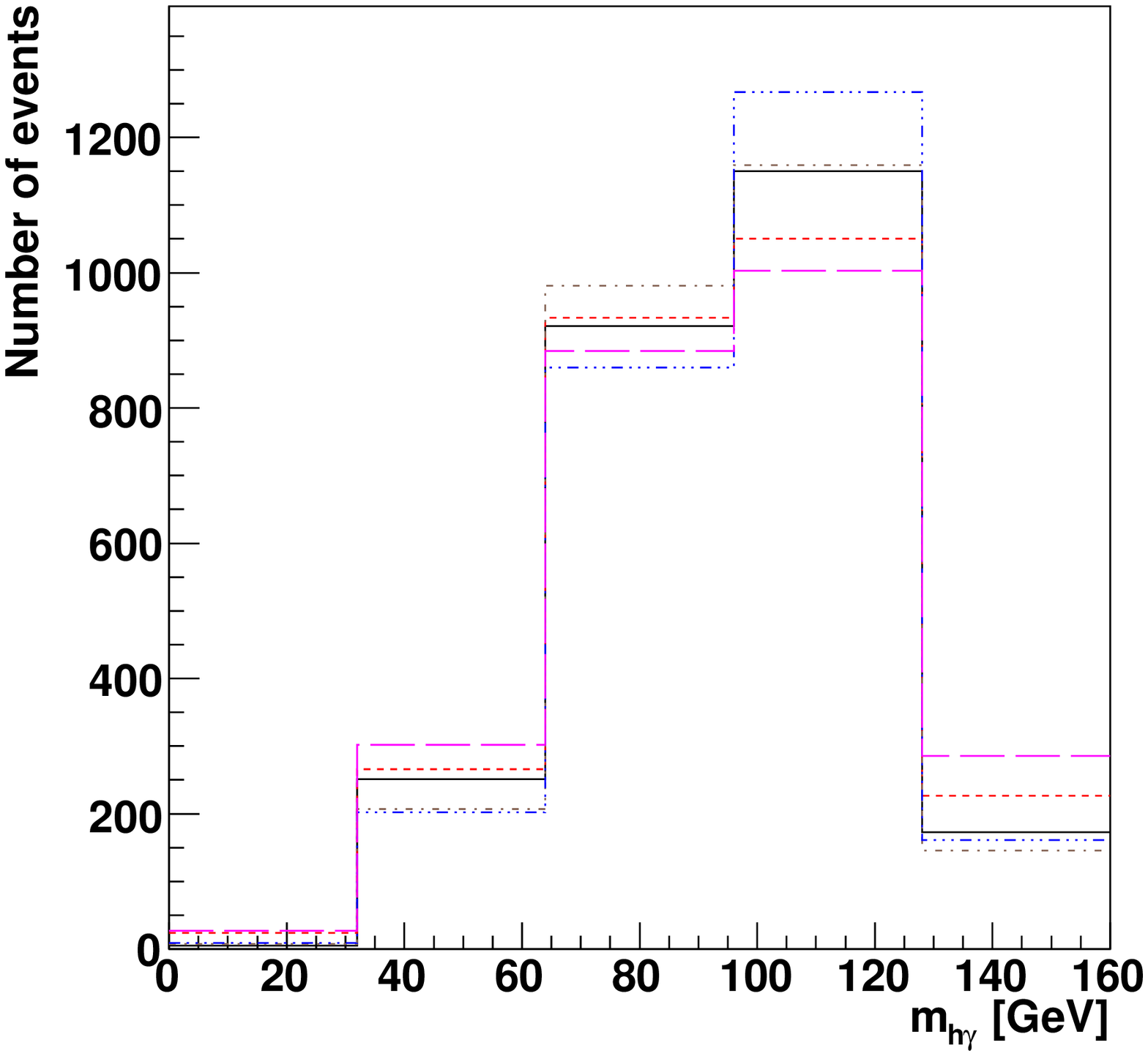,width=0.48\textwidth, bb=0 0 567 480, clip=true} &
    \epsfig{figure=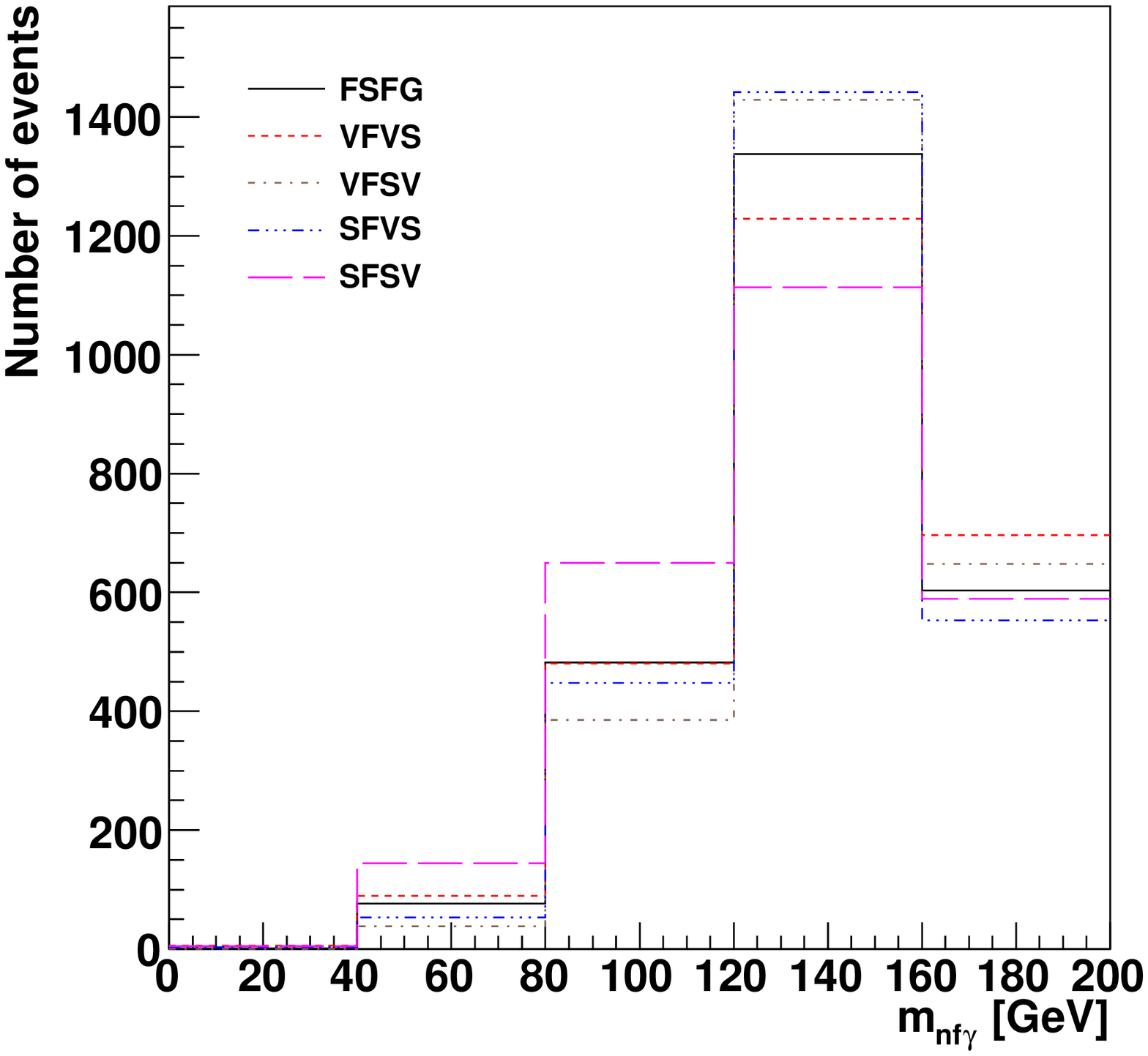,width=0.48\textwidth, bb=0 0 567 480, clip=true}
  \end{tabular}
  \mycaption{Reconstructed invariant mass distribution from ATLFAST detector 
    simulation, corresponding to G1a masses and cross sections for 10~fb$^{-1}$
    integrated luminosity. The histograms have been divided into 5 bins.}
  \label{fig:histsusy}
}

\FIGURE[ht]{
  \centering
  \begin{tabular}{cc} 
    \epsfig{figure=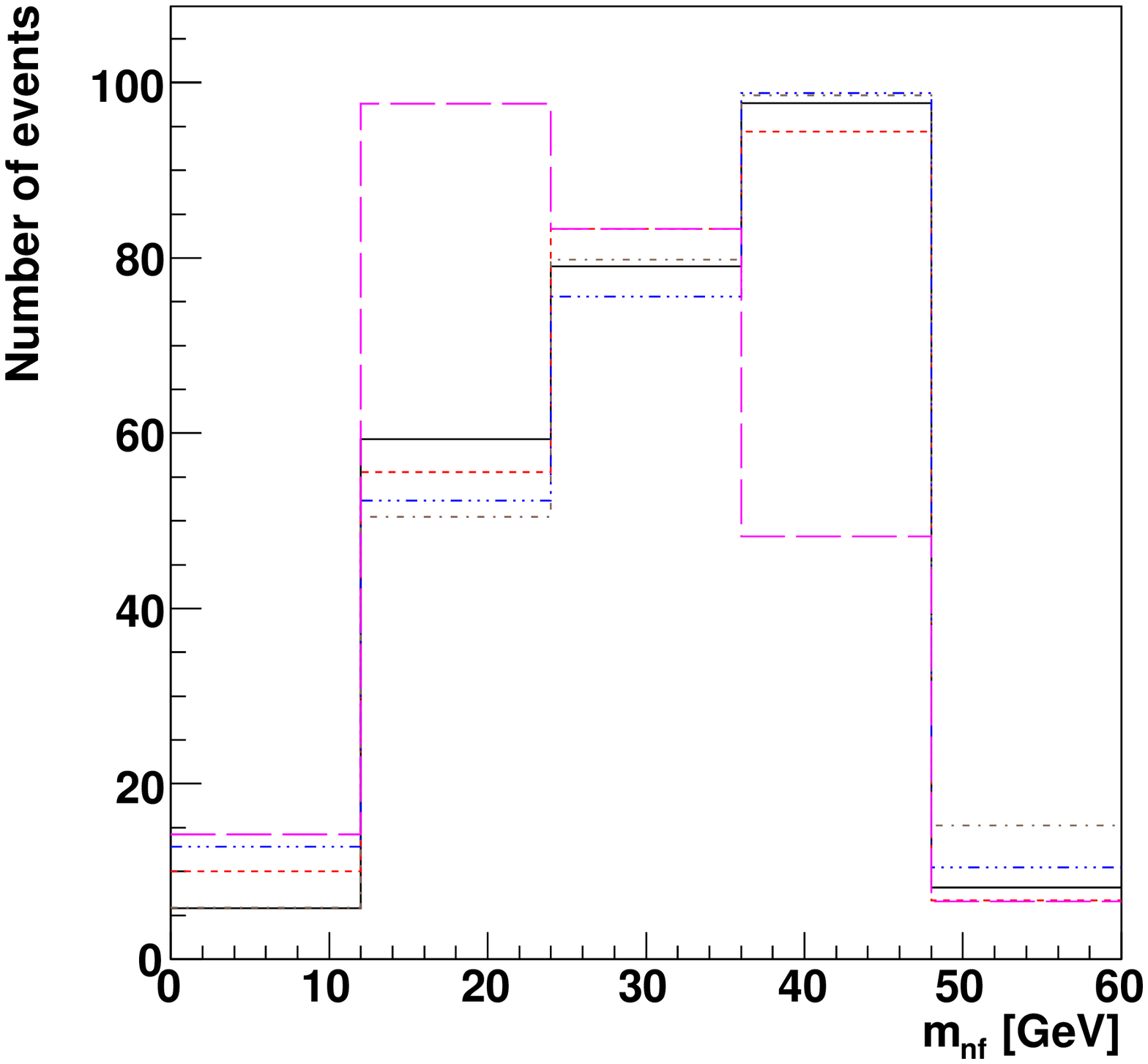,width=0.48\textwidth, bb=0 0 567 480, clip=true} &
    \epsfig{figure=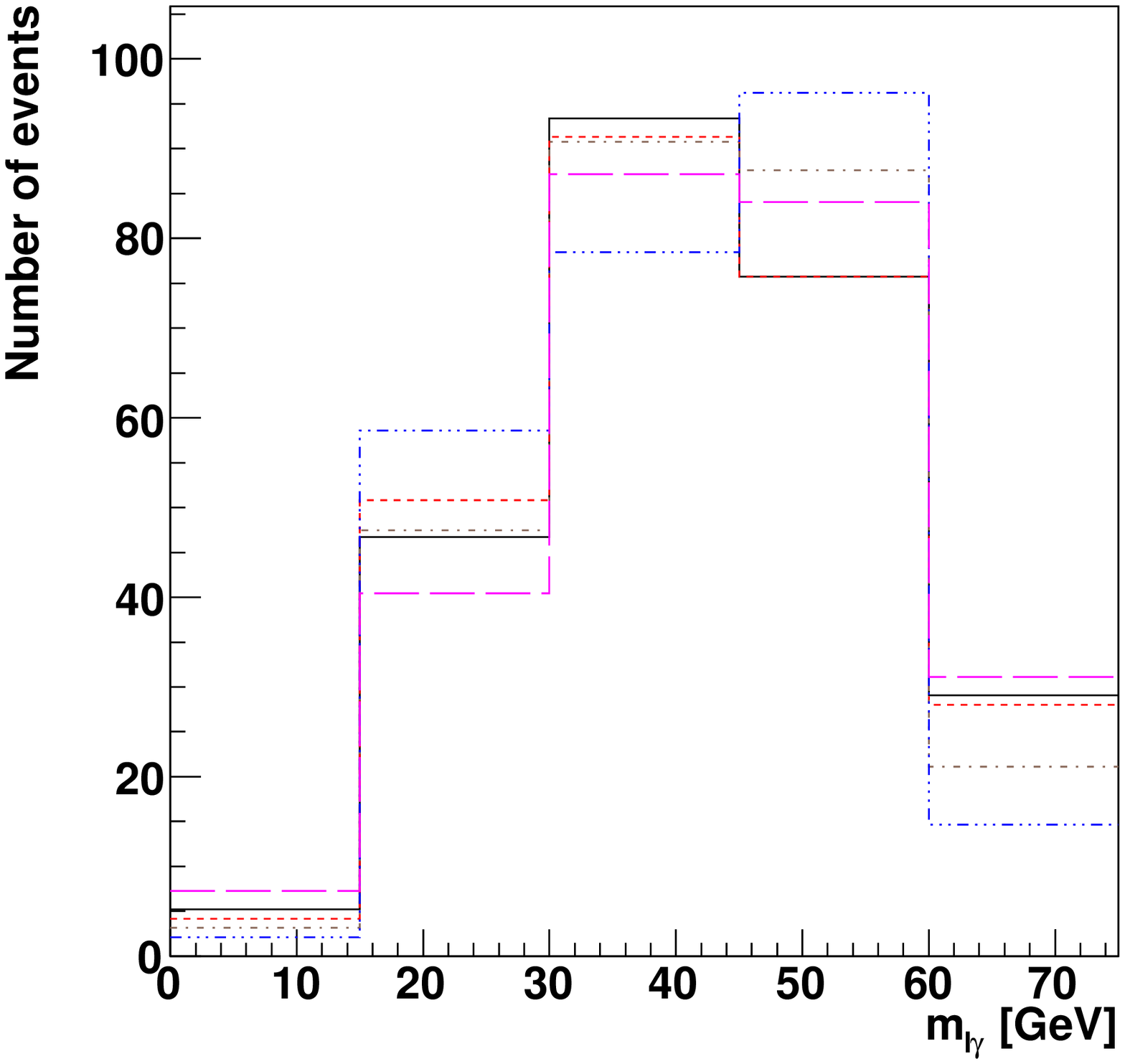,width=0.48\textwidth, bb=0 0 567 480, clip=true} \\
    \epsfig{figure=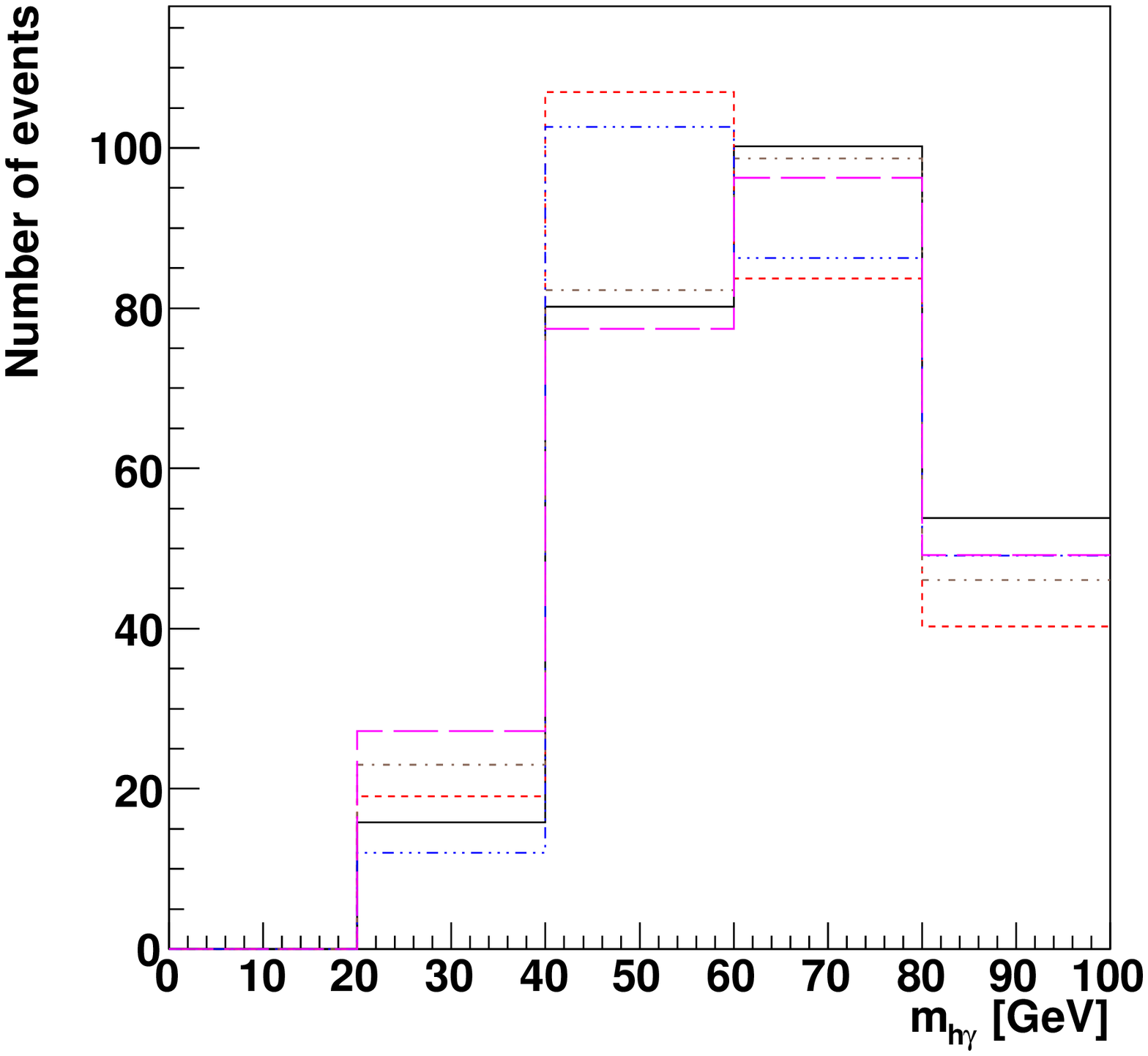,width=0.48\textwidth, bb=0 0 567 480, clip=true} &
    \epsfig{figure=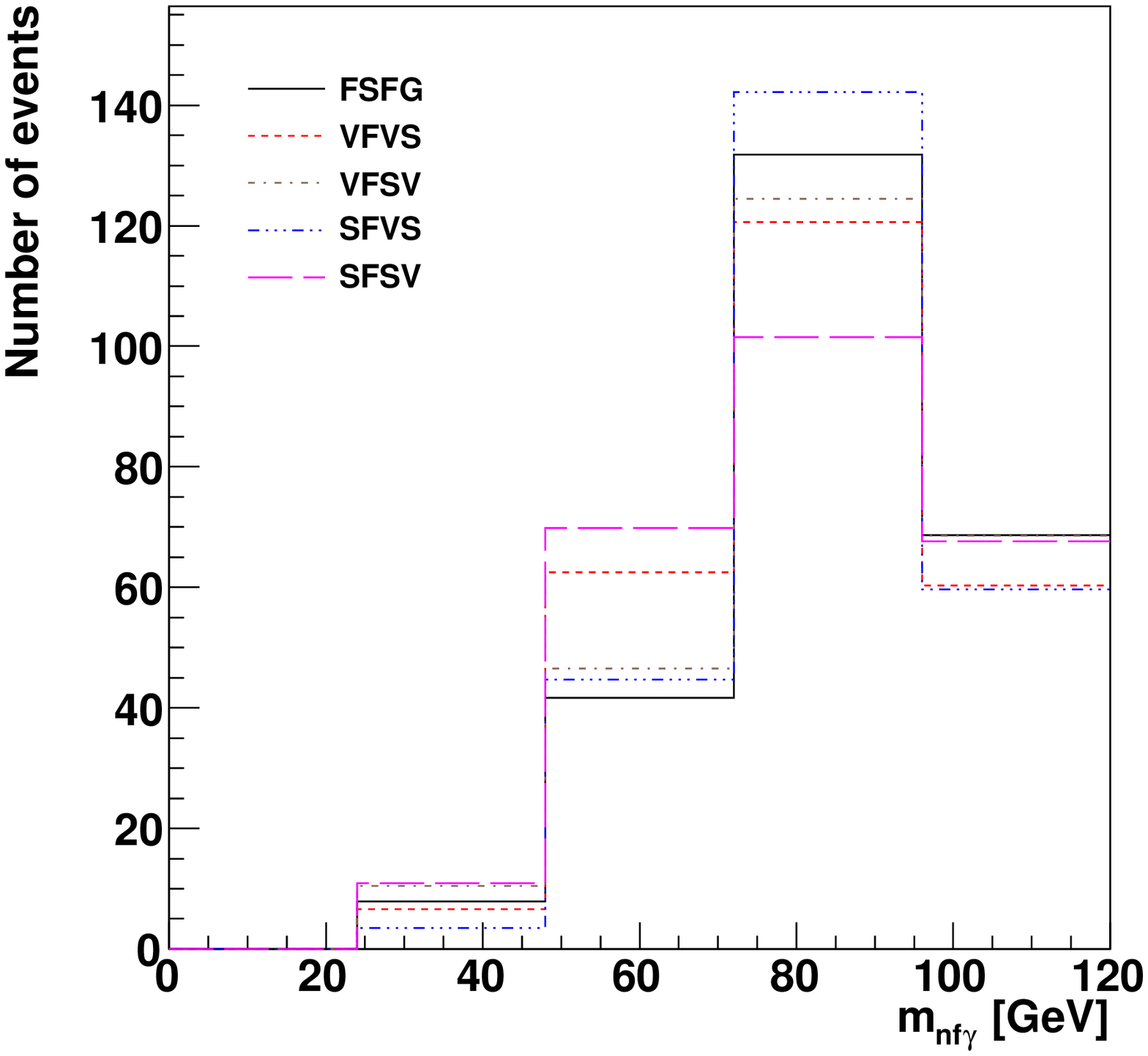,width=0.48\textwidth, bb=0 0 567 480, clip=true}
  \end{tabular}
  \mycaption{Same as Fig.~\ref{fig:histsusy}, but for U1 masses and cross
    sections.}
  \label{fig:histued}
}

\FIGURE[ht]{
  \centering
  \begin{tabular}{cc} 
    \epsfig{figure=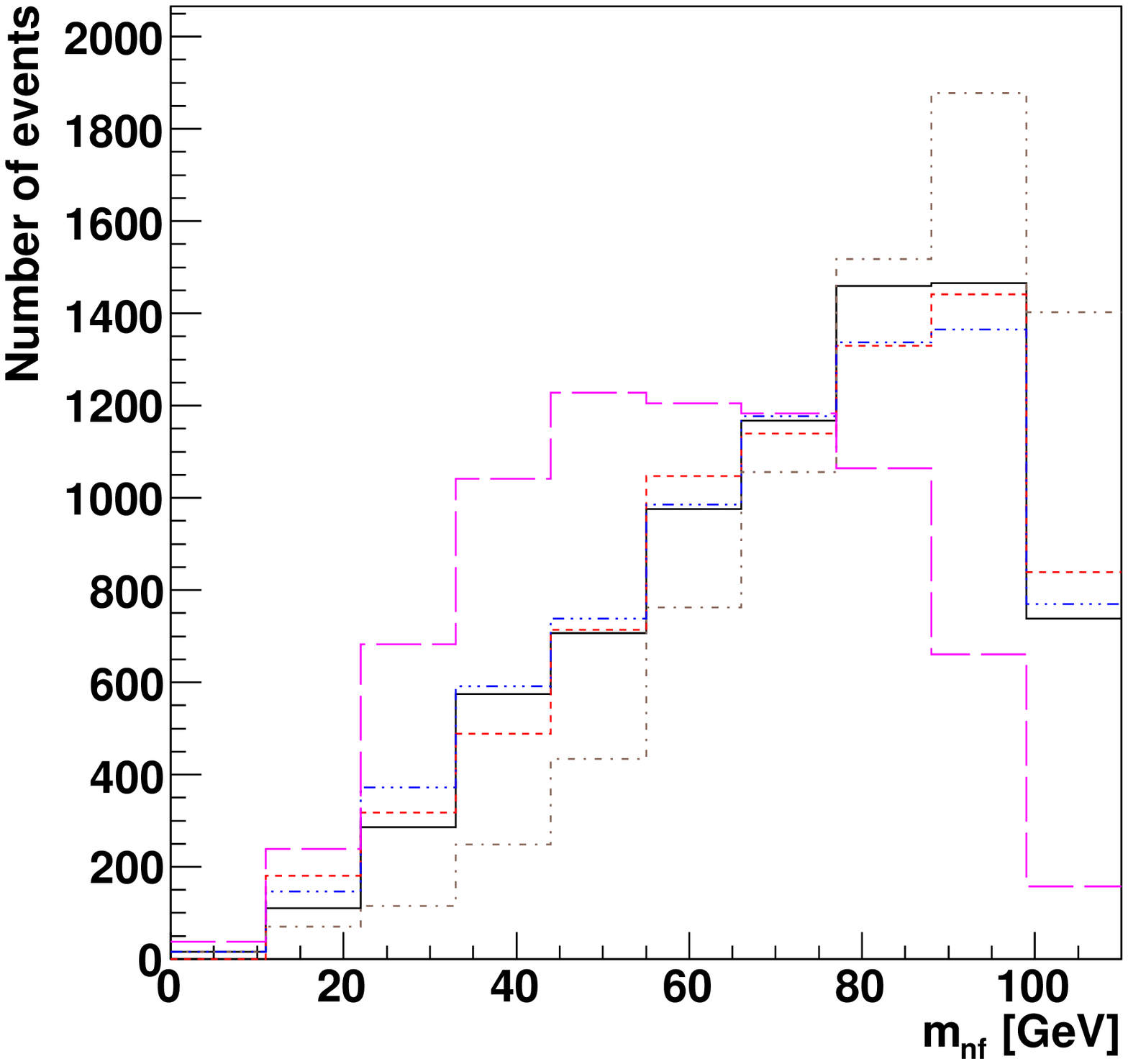,width=0.48\textwidth, bb=0 0 567 480, clip=true} &
    \epsfig{figure=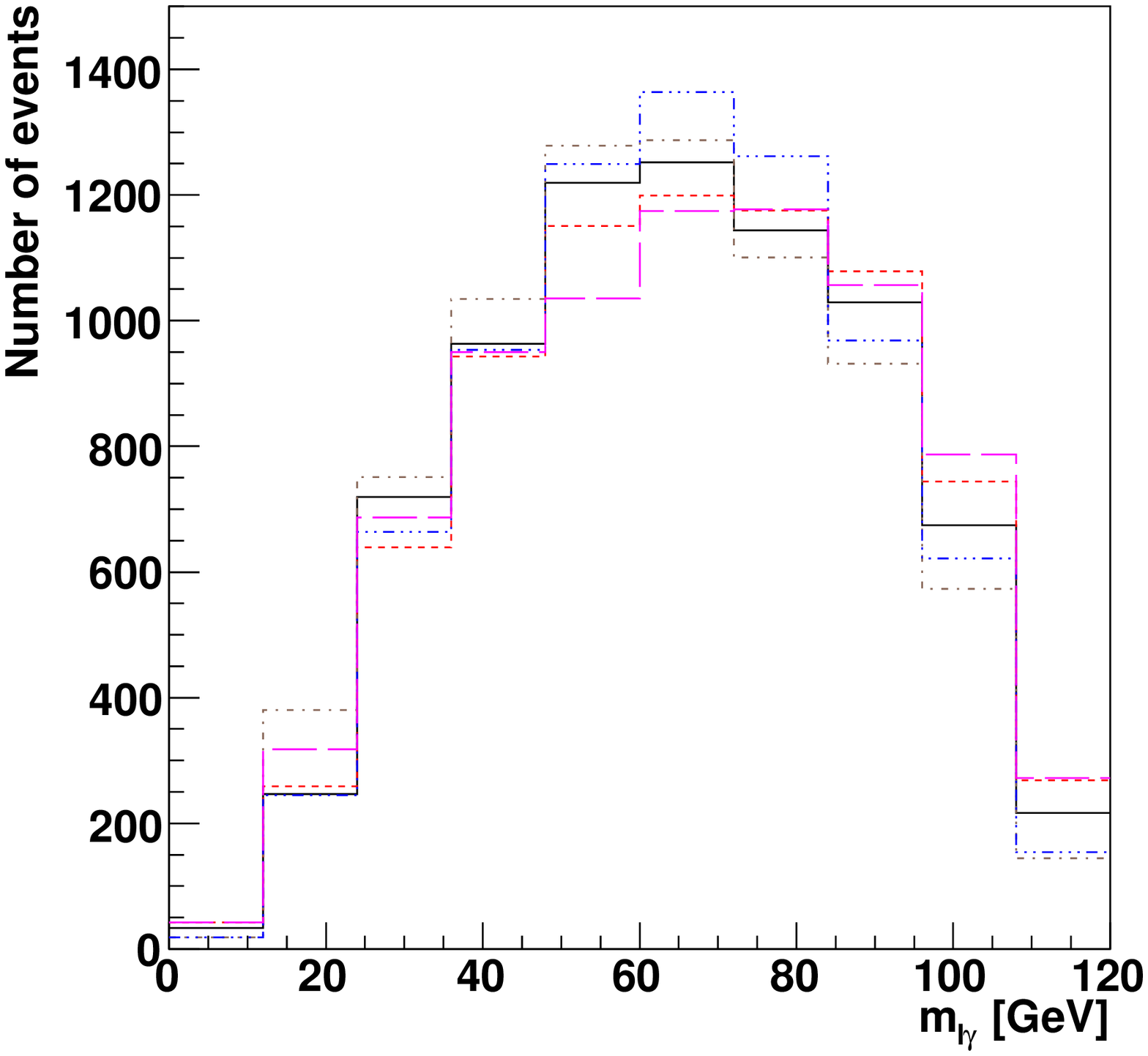,width=0.48\textwidth, bb=0 0 567 480, clip=true} \\
    \epsfig{figure=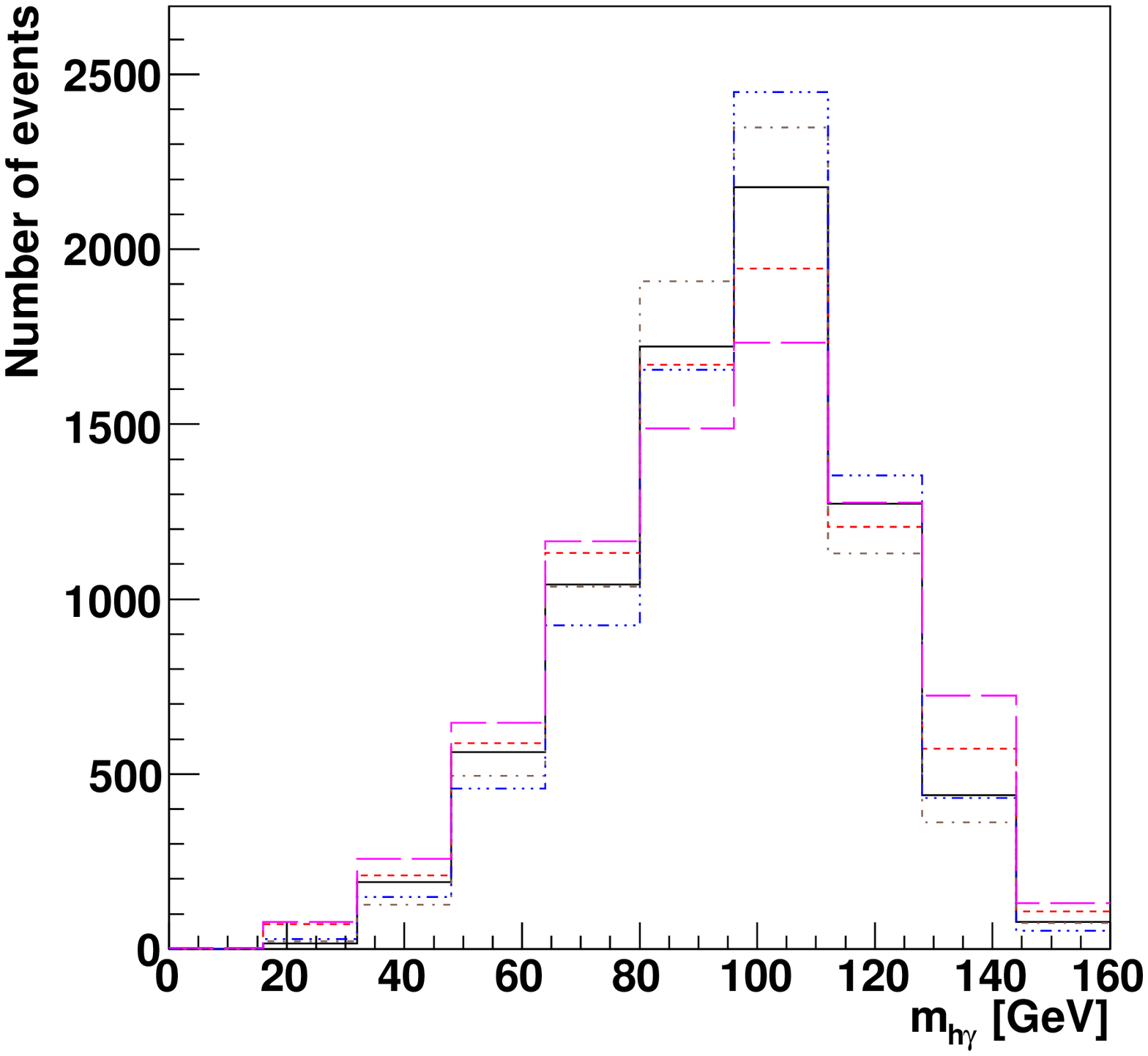,width=0.48\textwidth, bb=0 0 567 480, clip=true} &
    \epsfig{figure=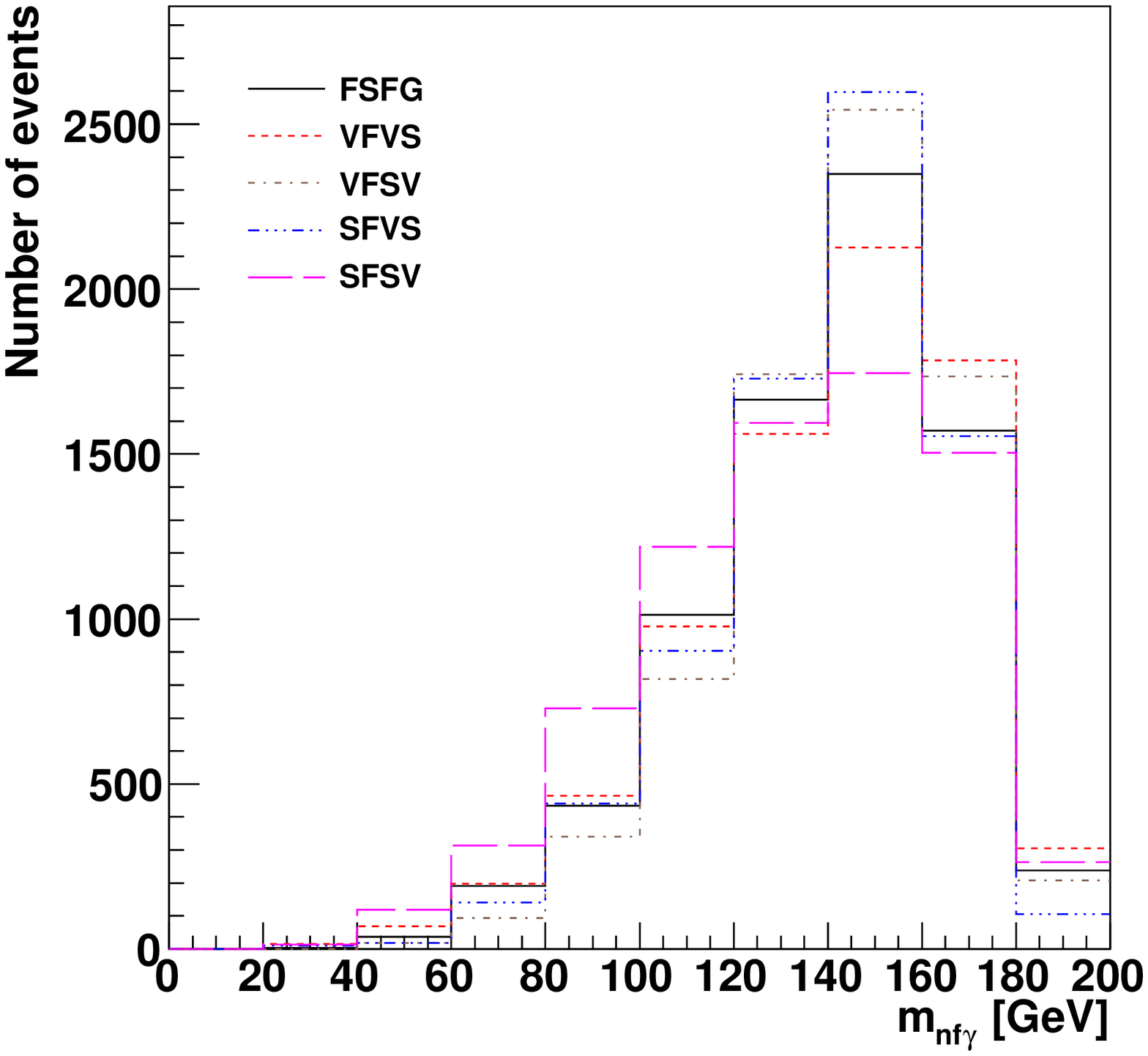,width=0.48\textwidth, bb=0 0 567 480, clip=true}
  \end{tabular}
  \mycaption{Reconstructed invariant mass distribution from ATLFAST detector 
    simulation, corresponding to G1a masses and cross sections for 30~fb$^{-1}$
    integrated luminosity. The histograms have been divided into 10 bins.}
  \label{fig:histsusy30}
}


Nevertheless, in summary the analysis of invariant mass distributions proves to
be a powerful tool to identify the spins of particles in decay chains with hard
photons. In this analysis 
the form the of the heavy particle interactions was fixed to the GMSB and UED6
predicitions.
In principle, the shapes of the invariant mass distributions are also 
sensitive to the couplings of those particles \cite{BKMP}, but a more
general analysis, where these parameters are taken as free variables, 
will be left for future work.


\section{Conclusion}
\label{sc:concl}

In this article we have analyzed the prospects for determining the spins of
new particles in decay chains with photons and missing energy at the LHC. As
concrete model incarnations of such signatures we considered supersymmetry with
gauge-mediated breaking (GMSB) and the standard model with two universal extra
dimensions (UED6). Each of these models predicts new partners of the SM
particles, the lightest of which is stable on grounds of a conserved parity. The
supersymmetric or KK partner of the photon can decay into this stable particle and
a hard photon. At
the LHC the partners of the colored particles (squarks/gluinos or
KK-quarks/KK-gluons) are produced with large cross sections and subsequently
could decay in several steps until the final photon emission step. As became well
known in recent years, the distributions of the invariant masses of two or more
of the decay products is sensitive to the spins of the decaying particles. The
measurement of the spins of the new particles is of central interest since this
is the crucial difference between supersymmetry and extra dimensions.

Following this approach, we first derived analytical expressions for the
invariant mass distributions for characteristic decay chains of those two
models. In greater detail we analyzed a class of decay chains that lead to two
leptons, one photon, and missing energy. This signature can stem from five
different decay processes (one in GMSB, four in UED6), which differ by the spins
of the intermediate particles. It was found that different invariant mass
distributions show different distinctive features between those five cases, 
so that the discriminative power is maximized by combining
information from all distributions. In a second step, we performed a realistic
phenomenological Monte-Carlo analysis including a fast detector simulation for
those five processes. Due to detector effects and cuts for background rejection, 
the reconstructed invariant mass distributions are distorted compared
to the analytical parton-level results. Nevertheless, the essential
characteristic features of the five different spin assignments are preserved, so
that the spins of the intermediate particles can be determined.

For a typical GMSB mass spectrum with superpartner masses below 1~TeV, we found
that with 10~fb$^{-1}$ integrated luminosity almost all of the five different
spin assignments, and in particular the GMSB and UED6 models, 
can be distinguished with 99.9\% confidence level. A typical
UED6 spectrum, on the other hand, is more degenerate, leading to smaller
branching ratios to leptons and to suppressed spin correlation effects in the
invariant mass distributions. As a result, for such a mass spectrum,
GMSB and UED6 can only be distinguished with a confidence level of less than
50\% with 10~fb$^{-1}$. A much higher luminosity would be needed for better
discrimination in this scenario.

Additional information could be obtained by looking at other decay
chains. We briefly investigated decay chains leading to a hard jet, a
photon, and missing energy, but no leptons. Such a decay chain has a large
branching fraction for UED6 scenarios and thus might be useful for discriminating
models in this case. A more conclusive answer would require a detailed analysis
of backgrounds to this process, which is beyond the scope of this work.

In summary, it appears feasible to distinguish GMSB and six-dimensional UED
models with LHC data alone if the mass spectrum is not very degenerate.

\acknowledgments

This work was supported in part by the Schweizer Nationalfonds. A.~F. 
is grateful for warm hospitality at the Universit\"at Z\"urich, Argonne National Laboratory and the
University of Chicago, where part of this work was performed.
W.~E. is grateful for warm hospitality at the Universit\"at Z\"urich,
where part of this work was performed.

We thank members of the ATLAS Collaboration for helpful discussions.
We have made use of the ATLAS physics analysis framework and tools
which are the result of collaboration-wide efforts.


\appendix

\section{Analytical results for invariant mass distributions}
\label{sc:app}

In the appendix we list analytical results for the invariant mass
distributions  of the various GMSB and UED6 decay chains discussed in
section~\ref{sc:spinc}. All those decay chains are of the form
\begin{equation}
D \rightarrow l_n^\pm C \rightarrow l_n^\pm l_f^\mp B \rightarrow l_n^\pm l_f^\mp
\gamma A\,,
\end{equation}
with $m_A < m_B < m_C < m_D$. Owing to kinematical constraints,
the invariant mass distributions are divided into
sections, which are bounded by the kinematical edges
\begin{align} 
(m_{f\gamma}^{\rm max})^2&=\frac{(m_C^2-m_B^2)(m_B^2-m_A^2)}{m_B^2}, \nonumber \\ 
(m_{n\gamma}^{\rm max})^2&=\frac{(m_D^2-m_C^2)(m_B^2-m_A^2)}{m_B^2}, \nonumber \\ 
(m_{No}^{\rm max})^2&=\frac{(m_D^2-m_C^2)(m_C^2-m_B^2)}{m_C^2}, \nonumber \\ 
(m_{n 2}^{\rm max})^2&=\frac{(m_D^2-m_C^2)(m_B^2-m_A^2)}{2 m_C^2-m_D^2}.
\label{eq:mmax}
\end{align}
Therefore the distributions take on the following structures:
\paragraph{The far-lepton--photon mass distribution \boldmath
 $\frac{{\rm d}P}{{\rm d}m_{f\gamma}^2}$:}
\begin{equation}
\frac{{\rm d}P}{{\rm d}m_{f\gamma}^2} = C1_f.
\end{equation}
\paragraph{The near-lepton--photon mass distribution \boldmath
 $\frac{{\rm d}P}{{\rm d}m_{n\gamma}^2}$:}
\begin{align}
\frac{{\rm d}P}{{\rm d}m_{n\gamma}^2} = \begin{cases}
C1_n\,, & 0 \leq m \leq m_{n\gamma}^{\rm max} \frac{m_B}{m_C}\\
C2_n\,, & m_{n\gamma}^{\rm max} \frac{m_B}{m_C} < m \leq m_{n\gamma}^{\rm max}.
\end{cases}
\end{align}
\paragraph{The high lepton--photon mass distribution \boldmath
 $\frac{{\rm d}P}{{\rm d}m_{h\gamma}^2}$:}
\subparagraph{Hierarchy A11: $m_{f\gamma}^{\rm max}<m_{n\gamma}^{\rm max} \frac{m_B}{m_C}<m_{n\gamma}^{\rm max}<m_{n 2}^{\rm max}$}
\begin{align}
\frac{{\rm d}P}{{\rm d}m_{h\gamma}^2} = \begin{cases}
C1_{hA11}\,, & 0 \leq m \leq m_{f\gamma}^{\rm max}\\
C2_{hA11}\,, & m_{f\gamma}^{\rm max} < m \leq m_{n\gamma}^{\rm max} \frac{m_B}{m_C} \\
C3_{hA11}\,, & m_{n\gamma}^{\rm max} \frac{m_B}{m_C} < m \leq m_{n 2}^{\rm max} . \end{cases}
\end{align}
\subparagraph{Hierarchy A12: $m_{n\gamma}^{\rm max} \frac{m_B}{m_C}<m_{f\gamma}^{\rm max}<m_{n\gamma}^{\rm max}<m_{n 2}^{\rm max}$}
\begin{align}
\frac{{\rm d}P}{{\rm d}m_{h\gamma}^2} = \begin{cases}
C1_{hA12}\,, & 0 \leq m \leq m_{n\gamma}^{\rm max}\frac{m_B}{m_C}\\
C2_{hA12}\,, & m_{n\gamma}^{\rm max}\frac{m_B}{m_C} < m \leq m_{f\gamma}^{\rm max} \\
C3_{hA12}\,, & m_{f\gamma}^{\rm max} < m \leq m_{n 2}^{\rm max} . \end{cases}
\end{align}
\subparagraph{Hierarchy A2: $m_{n\gamma}^{\rm max} \frac{mB}{mC} > m_{n2}^{\rm max} > m_{n\gamma}^{\rm max} > m_{f\gamma}^{\rm max}$}
\begin{align}
\frac{{\rm d}P}{{\rm d}m_{h\gamma}^2} = \begin{cases}
C1_{hA2}\,, & 0 \leq m \leq m_{n\gamma}^{\rm max} \frac{m_B}{m_C}\\
C2_{hA2}\,, & m_{n\gamma}^{\rm max} \frac{m_B}{m_C} < m \leq m_{n 2}^{\rm max} \\
C3_{hA2}\,, & m_{n 2}^{\rm max} < m \leq m_{f\gamma}^{\rm max} . \end{cases}
\end{align}
\subparagraph{Hierarchy B1: $m_{f\gamma}^{\rm max}<m_{n\gamma}^{\rm max} \frac{m_B}{m_C}<m_{n\gamma}^{\rm max}$}
\begin{align}
\frac{{\rm d}P}{{\rm d}m_{h\gamma}^2} = \begin{cases}
C1_{hB1}\,, & 0 \leq m \leq m_{f\gamma}^{\rm max}\\
C2_{hB1}\,, & m_{f\gamma}^{\rm max} < m \leq m_{n\gamma}^{\rm max} \frac{m_B}{m_C} \\
C3_{hB1}\,, & m_{n\gamma}^{\rm max} \frac{m_B}{m_C} < m \leq m_{n\gamma}^{\rm max} . \end{cases}
\end{align}
\subparagraph{Hierarchy B2: $m_{n\gamma}^{\rm max} \frac{m_B}{m_C}<m_{f\gamma}^{\rm max}<m_{n\gamma}^{\rm max}$}
\begin{align}
\frac{{\rm d}P}{{\rm d}m_{h\gamma}^2} = \begin{cases}
C1_{hB2}\,, & 0 \leq m \leq m_{n\gamma}^{\rm max} \frac{m_B}{m_C}\\
C2_{hB2}\,, & m_{n\gamma}^{\rm max} \frac{m_B}{m_C} < m \leq m_{f\gamma}^{\rm max} \\
C3_{hB2}\,, & m_{f\gamma}^{\rm max} < m \leq m_{n\gamma}^{\rm max} . \end{cases}
\end{align}
\paragraph{The low lepton-photon mass distribution \boldmath
 $\frac{{\rm d}P}{{\rm d}m_{l\gamma}^2}$:}
\subparagraph{Hierarchy A1: $m_{f\gamma}^{\rm max}<m_{n\gamma}^{\rm max}<m_{n 2}^{\rm max}$}
\begin{equation}
\frac{{\rm d}P}{{\rm d}m_{l\gamma}^2} = C1_{lA1}
\end{equation}
\subparagraph{Hierarchy A2: $m_{n 2}^{\rm max}<m_{n\gamma}^{\rm max}<m_{f\gamma}^{\rm max}$}
\begin{align}
\frac{{\rm d}P}{{\rm d}m_{l\gamma}^2} = \begin{cases}
C1_{lA2}\,, & 0 \leq m \leq m_{n 2}^{\rm max}\\
C2_{lA2}\,, & m_{n 2}^{\rm max} < m \leq m_{n\gamma}^{\rm max} .
\end{cases}
\end{align}
\subparagraph{Hierarchy B: $m_{f\gamma}^{\rm max}<m_{n\gamma}^{\rm max}$}
\begin{equation}
\frac{{\rm d}P}{{\rm d}m_{l\gamma}^2}=C1_{lB} .
\end{equation}
Many of these coefficients are related:
\begin{align}
&C1_f=C3_{A2} , \nonumber \\
&C1_n=C2_{hA11}=C2_{hB1} , \nonumber \\
&C2_n=C3_{hA11}=C3_{hA12}=C3_{hB1}=C3_{hB2}=C2_{lA2} , \nonumber \\
&C1_{hA11}=C1_{hA12}=C1_{hA2}=C1_{hB1}=C1_{hB2} , \nonumber \\
&C2_{hA12}=C2_{hA2}=C2_{hB2} , \nonumber \\
&C1_{lA1}=C1_{lA2}=C1_{lB} .
\end{align}
Below the results for the independent coefficients are given:
\paragraph{GMSB = phase space:}
\begin{align}
C1_f&=\frac{m_B^2}{\left(m_A^2-m_B^2\right) \left(m_B^2-m_C^2\right)}
\displaybreak[0] \,, \\[1.5ex]
C1_n&=\frac{m_B^2 m_C^2 \mathrm{log}\left[\frac{m_B^2}{m_C^2}\right]}{\left(m_A^2-m_B^2\right) \left(m_B^2-m_C^2\right) \left(m_C^2-m_D^2\right)}
\displaybreak[0] \,, \\[1.5ex]
C2_n&=\frac{m_B^2 m_C^2 \mathrm{log}\left[\frac{m^2 m_B^2}{\left(m_A^2-m_B^2\right) \left(m_C^2-m_D^2\right)}\right]}{\left(m_A^2-m_B^2\right) \left(m_B^2-m_C^2\right) \left(m_C^2-m_D^2\right)}
\displaybreak[0] \,, \\[1.5ex]
C1_{hA11}&=-\frac{m_B^2 m_C^2 \left(m^2+2 \left(m^2-m_A^2+m_B^2\right) \mathrm{arccoth}\left[\frac{m^2-2 m_A^2+2 m_B^2}{m^2}\right]\right)}{\left(m_A^2-m_B^2\right) \left(m^2-m_A^2+m_B^2\right) \left(m_B^2-m_C^2\right) \left(m_C^2-m_D^2\right)}
\displaybreak[0] \,, \\[1.5ex]
C2_{hA12}&=-\frac{m_B^2 m_C^2 \left(m^2+\left(m^2-m_A^2+m_B^2\right) \mathrm{log}\left[-\frac{\left(m^2-m_A^2+m_B^2\right) \left(m_C^2-m_D^2\right)}{m^2 m_C^2}\right]\right)}{\left(m_A^2-m_B^2\right) \left(m^2-m_A^2+m_B^2\right) \left(m_B^2-m_C^2\right) \left(m_C^2-m_D^2\right)}
\displaybreak[0] \,, \\[1.5ex]
C1_{lA1}&=\Bigg(m_B^2 m_C^2 \bigg(2 m^2-m_A^2+m_B^2-\frac{(m^2-m_A^2+m_B^2) m_D^2}{m_C^2}\nonumber \\
&\qquad +(m^2-m_A^2+m_B^2) \mathrm{log}\left[\frac{m_B^2 (m^2-m_A^2+m_B^2)}{(-m_A^2+m_B^2) m_C^2}\right]\bigg)\Bigg)\nonumber \\
&\qquad /\Big((m_A^2-m_B^2) (m^2-m_A^2+m_B^2) (m_B^2-m_C^2)
(m_C^2-m_D^2)\Big)
\,.
\end{align}
\allowdisplaybreaks
\paragraph{VFVS:}
\begin{align}
C1_f&=\Big(3 m_B^4 (2 m^4 m_B^2-2 m^2 (m_A^2-m_B^2) (m_B^2-m_C^2)+(m_A^2-m_B^2)^2\nonumber \\
&\qquad \times (m_B^2-m_C^2))\Big)/\Big((m_A^2-m_B^2)^3 (m_B^2-m_C^2)^2 (2 m_B^2+m_C^2)\Big)
\displaybreak[0] \,, \\[1.5ex]
C1_n&=\Big(6 m_B^2 m_C^2 ((m_B^2-m_C^2) (4 m^2 m_B^2 m_C^2 (m_C^2-2 m_D^2) \nonumber \\
&\qquad+(m_A^2-m_B^2) (m_C^2-m_D^2) (m_B^2 m_C^2-2 (2 m_B^2+m_C^2) m_D^2)) \nonumber \\
&\qquad-2 m_B^2 (2 m^2 m_C^2 (m_B^2+m_C^2) (m_C^2-2 m_D^2) \nonumber \\
&\qquad+(m_A^2-m_B^2) (m_C^2-m_D^2) (m_C^4-2 (m_B^2+2 m_C^2) m_D^2)) \mathrm{log}\left[\frac{m_B}{m_C}\right])\Big) \nonumber \\
&\qquad/\Big((m_A^2-m_B^2)^2 (m_B^2-m_C^2)^2 (2 m_B^2+m_C^2) (m_C^2-m_D^2)^2 (m_C^2+2 m_D^2)\Big)
\displaybreak[0] \,, \\[1.5ex]
C2_n&=\Bigg(6 m_B^2 m_C^2 \bigg(m_C^2 (2 m^4 m_B^4 m_C^2-(m_A^2-m_B^2)^2 (m_C^2-m_D^2)  \nonumber \\
&\qquad\times(3 m_B^2 m_C^2-2 (5 m_B^2+m_C^2) m_D^2)+m^2 m_B^2 (m_A^2-m_B^2) \nonumber \\
&\qquad\times(-2 m_C^4+m_B^2 (3 m_C^2-10 m_D^2)))+m_B^2 (m_A^2-m_B^2) (2 m^2 m_C^2 (m_B^2+m_C^2) \nonumber \\
&\qquad\times(m_C^2-2 m_D^2)+(m_A^2-m_B^2) (m_C^2-m_D^2) (m_C^4-2 (m_B^2+2 m_C^2) m_D^2)) \nonumber \\
&\qquad\times\mathrm{log}\left[\frac{(m_A^2-m_B^2) (m_C^2-m_D^2)}{m^2 m_B^2}\right]\bigg)\Bigg) \nonumber \\
&\qquad/\Big((m_A^2-m_B^2)^3 (m_B^2-m_C^2)^2 (2 m_B^2+m_C^2) (m_C^2-m_D^2)^2 (m_C^2+2 m_D^2)\Big)
\displaybreak[0] \,, \\[1.5ex]
C1_{hA11}&=\Bigg(3 m_B^4 m_C^2 \bigg(m^2 (-2 (m_A^2-m_B^2)^4 (m_C^2-m_D^2) \nonumber \\
&\qquad\times (m_C^4+2 m_B^2 m_D^2-8 m_C^2 m_D^2)+6 m^8 m_B^2 (m_C^4-4 m_C^2 m_D^2+2 m_D^4) \nonumber \\
&\qquad-2 m^6 (m_A^2-m_B^2) (m_C^6+6 m_C^4 m_D^2-8 m_C^2 m_D^4\nonumber \\
&\qquad +9 m_B^2 (m_C^4-4 m_C^2 m_D^2+2 m_D^4))+m^4 (m_A^2-m_B^2)^2\nonumber \\
&\qquad \times (5 m_C^6+34 m_C^4 m_D^2-44 m_C^2 m_D^4+m_B^2 (19 m_C^4-78 m_C^2 m_D^2+40 m_D^4))\nonumber \\
&\qquad -m^2 (m_A^2-m_B^2)^3(m_B^2 (7 m_C^4-34 m_C^2 m_D^2+20 m_D^4)\nonumber \\
&\qquad -m_C^2 (m_C^4-44 m_C^2 m_D^2+44 m_D^4)))-2 (m_A^2-m_B^2) (m^2-m_A^2+m_B^2)^3\nonumber \\
&\qquad \times (2 m^2 m_C^2 (m_B^2+m_C^2) (m_C^2-2 m_D^2)+(m_A^2-m_B^2) (m_C^2-m_D^2)\nonumber \\
&\qquad \times (m_C^4-2 (m_B^2+2 m_C^2) m_D^2))\mathrm{log}\left[1-\frac{m^2}{m^2-m_A^2+m_B^2}\right]\bigg)\Bigg)\nonumber \\
&\qquad /\Big((m_A^2-m_B^2)^3 (m^2-m_A^2+m_B^2)^3(m_B^2-m_C^2)^2\nonumber \\
&\qquad \times (2 m_B^2+m_C^2) (m_C^2-m_D^2)^2 (m_C^2+2 m_D^2)\Big)
\displaybreak[0] \,, \\[1.5ex]
C2_{hA12}&=\Bigg(3 m_B^4 m_C^2 \bigg(\frac{1}{(m^2-m_A^2+m_B^2)^3}(2 (m_A^2-m_B^2)^5 (m_B^2+2 m_C^2)\nonumber \\
&\qquad \times (m_C^4-5 m_C^2 m_D^2+4 m_D^4)+2 m^{10} m_B^2 (5 m_C^4-12 m_C^2 m_D^2+6 m_D^4)\nonumber \\
&\qquad-m^4 (m_A^2-m_B^2)^3 (-25 m_C^6+152 m_C^4 m_D^2-92 m_C^2 m_D^4\nonumber \\
&\qquad+m_B^2 (11 m_C^4+8 m_C^2 m_D^2-4 m_D^4))-2 m^8 (m_A^2-m_B^2) \nonumber \\
&\qquad \times (-m_C^2 (m_C^4-14 m_C^2 m_D^2+8 m_D^4)+2 m_B^2 (8 m_C^4-17 m_C^2 m_D^2+9 m_D^4))\nonumber \\
&\qquad-2 m^2 (m_A^2-m_B^2)^4 (9 m_C^6-47 m_C^4 m_D^2+32 m_C^2 m_D^4\nonumber \\
&\qquad+m_B^2 (2 m_C^4-15 m_C^2 m_D^2+10 m_D^4))+m^6 (m_A^2-m_B^2)^2\nonumber \\
&\qquad \times (-11 m_C^6+102 m_C^4 m_D^2-60 m_C^2 m_D^4+m_B^2 (35 m_C^4-56 m_C^2 m_D^2+32 m_D^4)))\nonumber \\
&\qquad+2 (m_A^2-m_B^2) (2 m^2 m_C^2 (m_B^2+m_C^2) (m_C^2-2 m_D^2)+(m_A^2-m_B^2)\nonumber \\
&\qquad \times (m_C^2-m_D^2) (m_C^4-2 (m_B^2+2 m_C^2) m_D^2)) \nonumber \\
&\qquad \times \mathrm{log}\left[\frac{(m^2-m_A^2+m_B^2) (-m_C^2+m_D^2)}{m^2 m_C^2}\right]\bigg)\Bigg)\nonumber \\
&\qquad/\Big((m_A^2-m_B^2)^3 (m_B^2-m_C^2)^2 (2 m_B^2+m_C^2) (m_C^2-m_D^2)^2 (m_C^2+2 m_D^2)\Big)
\displaybreak[0] \,, \\[1.5ex]
C1_{lA1}&=\Bigg(3 m_B^4 m_C^2 \bigg(\frac{1}{m_B^2 m_C^2}(2 m^{10} m_B^4 (-2 m_C^6+12 m_C^4 m_D^2-9 m_C^2 m_D^4+2 m_D^6)\nonumber \\
&\qquad+(m_A^2-m_B^2)^5 (m_B^2-m_C^2) (m_C^2-m_D^2) (4 m_C^4 m_D^2\nonumber \\
&\qquad+m_B^2 (-3 m_C^4+7 m_C^2 m_D^2+2 m_D^4))+2 m^8 m_B^2 (m_A^2-m_B^2)\nonumber \\
&\qquad \times (-2 m_C^8+14 m_C^6 m_D^2-11 m_C^4 m_D^4+2 m_C^2 m_D^6\nonumber \\
&\qquad+m_B^2 (9 m_C^6-44 m_C^4 m_D^2+30 m_C^2 m_D^4-8 m_D^6))+m^2 (m_A^2-m_B^2)^4\nonumber \\
&\qquad \times (12 m_C^6 m_D^2 (m_C^2-m_D^2)-m_B^2 m_C^2 (m_C^6+16 m_C^4 m_D^2-19 m_C^2 m_D^4+10 m_D^6)\nonumber \\
&\qquad+m_B^4 (3 m_C^6-10 m_C^4 m_D^2+5 m_C^2 m_D^4+10 m_D^6))-m^6 (m_A^2-m_B^2)^2\nonumber \\
&\qquad (-4 m_C^8 m_D^2+4 m_C^6 m_D^4+m_B^4 (28 m_C^6-116 m_C^4 m_D^2+71 m_C^2 m_D^4-26 m_D^6)\nonumber \\
&\qquad+m_B^2 m_C^2 (-10 m_C^6+76 m_C^4 m_D^2-61 m_C^2 m_D^4+14 m_D^6))+m^4 (m_A^2-m_B^2)^3\nonumber \\
&\qquad \times (12 m_C^6 m_D^2 (-m_C^2+m_D^2)+m_B^4 (14 m_C^6-52 m_C^4 m_D^2+29 m_C^2 m_D^4-22 m_D^6)\nonumber \\
&\qquad+m_B^2 m_C^2 (-10 m_C^6+74 m_C^4 m_D^2-59 m_C^2 m_D^4+18 m_D^6)))\nonumber \\
&\qquad+2 (m_A^2-m_B^2) (m^2-m_A^2+m_B^2)^3 (2 m^2 m_C^2 (m_B^2+m_C^2) (m_C^2-2 m_D^2) \nonumber \\
&\qquad +(m_A^2-m_B^2) (m_C^2-m_D^2) (m_C^4-2 (m_B^2+2 m_C^2) m_D^2))\nonumber \\
&\qquad \times \mathrm{log}\left[\frac{(m_B^2-m_A^2) m_C^2}{m_B^2(m^2-m_A^2+m_B^2)}\right]\bigg)\Bigg)/\Big((m_A^2-m_B^2)^3 (m^2-m_A^2+m_B^2)^3\nonumber \\
&\qquad (m_B^2-m_C^2)^2 (2 m_B^2+m_C^2) (m_C^2-m_D^2)^2 (m_C^2+2 m_D^2)\Big)
\,.
\end{align}
\paragraph{VFSV:}
\begin{align}
C1_f&=\frac{m_B^2}{\left(-m_A^2+m_B^2\right) \left(-m_B^2+m_C^2\right)}
\displaybreak[0] \,, \\[1.5ex]
C1_n&=\frac{2 m_B^2 m_C^2 \left(-\left(m_B^2-m_C^2\right) \left(m_C^2-2 m_D^2\right)+2 m_C^2 \left(m_B^2-2 m_D^2\right) \mathrm{log}\left[\frac{m_B}{m_C}\right]\right)}{\left(m_A^2-m_B^2\right) \left(m_B^2-m_C^2\right)^2 \left(m_C^4+m_C^2 m_D^2-2 m_D^4\right)}
\displaybreak[0] \,, \\[1.5ex]
C2_n&=-\Bigg(2 m_B^2 m_C^4 \bigg((m_C^2-2 m_D^2) (m^2 m_B^2-(m_A^2-m_B^2) (m_C^2-m_D^2))\nonumber \\
&\qquad +(m_A^2-m_B^2) (m_B^2-2 m_D^2) (-m_C^2+m_D^2)\nonumber \\
&\qquad \times \mathrm{log}\left[\frac{m^2 m_B^2}{(m_A^2-m_B^2) (m_C^2-m_D^2)}\right]\bigg)\Bigg)\nonumber \\
&\qquad /\left(\left(m_A^2-m_B^2\right)^2 \left(m_B^2-m_C^2\right)^2 \left(m_C^2-m_D^2\right)^2 \left(m_C^2+2 m_D^2\right)\right)
\displaybreak[0] \,, \\[1.5ex]
C1_{hA11}&=\Bigg(m_B^2 m_C^4 \bigg(m^2 (2 m^6 m_B^2 (m_C^2-2 m_D^2)-2 (m_A^2-m_B^2)^3 (m_C^2-m_D^2)\nonumber \\
&\qquad \times (m_B^2-2 m_C^2+2 m_D^2)-m^4 (m_A^2-m_B^2) (-3 m_C^4+4 m_C^2 m_D^2\nonumber \\
&\qquad +m_B^2 (7 m_C^2-12 m_D^2))+m^2 (m_A^2-m_B^2)^2 (-9 m_C^4+16 m_C^2 m_D^2-4 m_D^4\nonumber \\
&\qquad +m_B^2 (7 m_C^2-10 m_D^2)))-4 (m_A^2-m_B^2) (m^2-m_A^2+m_B^2)^3 \nonumber \\
&\qquad \times (m_B^2-2 m_D^2) (m_C^2-m_D^2) \mathrm{arccoth}\left[\frac{m^2-2 m_A^2+2 m_B^2}{m^2}\right]\bigg)\Bigg)\nonumber \\
&\qquad /\Big((m_A^2-m_B^2)^2 (m^2-m_A^2+m_B^2)^3 (m_B^2-m_C^2)^2\nonumber \\
&\qquad \times (m_C^2-m_D^2)^2 (m_C^2+2 m_D^2)\Big)
\displaybreak[0] \,, \\[1.5ex]
C2_{hA12}&=\Bigg(2 m_B^2 m_C^4 \bigg(\frac{1}{2 (m^2-m_A^2+m_B^2)^3 m_C^2}(2 m_B^2 (-m_A^2+m_B^2)^3\nonumber \\
&\qquad \times (m_C^4-3 m_C^2 m_D^2+2 m_D^4)+m^6 (3 m_C^6-4 m_C^4 m_D^2\nonumber \\
&\qquad +m_B^2 (m_C^4-6 m_C^2 m_D^2+4 m_D^4))+2 m^2 (m_A^2-m_B^2)^2 (2 (m_C^3-m_C m_D^2)^2\nonumber \\
&\qquad +m_B^2 (3 m_C^4-10 m_C^2 m_D^2+6 m_D^4))-m^4 (m_A^2-m_B^2)\nonumber \\
&\qquad \times (9 m_C^6-16 m_C^4 m_D^2+4 m_C^2 m_D^4+m_B^2 (5 m_C^4-20 m_C^2 m_D^2+12 m_D^4)))\nonumber \\
&\qquad +(m_B^2-2 m_D^2) (-m_C^2+m_D^2) \mathrm{log}\left[-\frac{(m^2-m_A^2+m_B^2) (m_C^2-m_D^2)}{m^2 m_C^2}\right]\bigg)\Bigg)\nonumber \\
&\qquad /\left(\left(m_A^2-m_B^2\right) \left(m_B^2-m_C^2\right)^2 \left(m_C^2-m_D^2\right)^2 \left(m_C^2+2 m_D^2\right)\right)
\displaybreak[0] \,, \\[1.5ex]
C1_{lA1}&=\Bigg(m_B^2 \bigg(-2 m^8 m_B^2 m_C^4 (m_C^2-2 m_D^2)+(m_A^2-m_B^2)^4 (m_B^2-m_C^2)\nonumber \\
&\qquad \times (m_C^6-6 m_C^4 m_D^2+7 m_C^2 m_D^4-2 m_D^6)-m^2 (m_A^2-m_B^2)^3 (m_C^2-m_D^2)\nonumber \\
&\qquad \times (m_C^6+11 m_C^4 m_D^2-6 m_C^2 m_D^4+m_B^2 (m_C^4-15 m_C^2 m_D^2+6 m_D^4))\nonumber \\
&\qquad +m^6 (m_A^2-m_B^2) (-2 m_C^8-2 m_C^6 m_D^2+7 m_C^4 m_D^4-2 m_C^2 m_D^6\nonumber \\
&\qquad +m_B^2 (6 m_C^6-6 m_C^4 m_D^2-7 m_C^2 m_D^4+2 m_D^6))-m^4 (m_A^2-m_B^2)^2\nonumber \\
&\qquad \times (-6 m_C^8-2 m_C^6 m_D^2+17 m_C^4 m_D^4-6 m_C^2 m_D^6\nonumber \\
&\qquad +m_B^2 (4 m_C^6+8 m_C^4 m_D^2-21 m_C^2 m_D^4+6 m_D^6))\nonumber \\
&\qquad -2 (m_A^2-m_B^2)(m^2-m_A^2+m_B^2)^3 m_C^4 (m_B^2-2 m_D^2) (m_C^2-m_D^2) \nonumber \\
&\qquad \times \mathrm{log}\left[\frac{(-m_A^2+m_B^2) m_C^2}{m_B^2 (m^2-m_A^2+m_B^2)}\right]\bigg)\Bigg)/\Big((m_A^2-m_B^2)^2 (m^2-m_A^2+m_B^2)^3\nonumber \\
&\qquad \times (m_B^2-m_C^2)^2 (m_C^2-m_D^2)^2 (m_C^2+2 m_D^2)\Big)
\,.
\end{align}
\paragraph{SFVS:}
\begin{align}
C1_f&=\Big(3 m_B^4 (2 m^4 m_B^2-2 m^2 (m_A^2-m_B^2) (m_B^2-m_C^2)+(m_A^2-m_B^2)^2\nonumber \\
&\qquad \times (m_B^2-m_C^2))\Big)/\Big((m_A^2-m_B^2)^3 (m_B^2-m_C^2)^2 (2 m_B^2+m_C^2)\Big)
\displaybreak[0] \,, \\[1.5ex]
C1_n&=-\Bigg(6 m_B^4 m_C^2 \bigg((m_B^2-m_C^2) (-4 m^2 m_C^2-(m_A^2-m_B^2) (m_C^2-m_D^2))\nonumber \\
&\qquad +2 m_C^2 (2 m^2 (m_B^2+m_C^2)+(m_A^2-m_B^2) (m_C^2-m_D^2)) \mathrm{log}\left[\frac{m_B}{m_C}\right]\bigg)\Bigg)\nonumber \\
&\qquad /\Big((m_A^2-m_B^2)^2 (m_B^2-m_C^2)^2 (2 m_B^2+m_C^2) (m_C^2-m_D^2)^2\Big)
\displaybreak[0] \,, \\[1.5ex]
C2_n&=\Bigg(6 m_B^4 m_C^4 \bigg(2 m^4 m_B^2 m_C^2+m^2 (m_A^2-m_B^2) (3 m_B^2-2 m_C^2) (m_C^2-m_D^2)\nonumber \\
&\qquad -3 (m_A^2-m_B^2)^2 (m_C^2-m_D^2)^2-(m_A^2-m_B^2) (m_C^2-m_D^2)\nonumber \\
&\qquad \times (2 m^2 (m_B^2+m_C^2)+(m_A^2-m_B^2) (m_C^2-m_D^2))\nonumber \\
&\qquad \times \mathrm{log}\left[\frac{m^2 m_B^2}{(m_A^2-m_B^2) (m_C^2-m_D^2)}\right]\bigg)\Bigg)\nonumber \\
&\qquad /\Big((m_A^2-m_B^2)^3 (m_B^2-m_C^2)^2 (2 m_B^2+m_C^2) (m_C^2-m_D^2)^3\Big)
\displaybreak[0] \,, \\[1.5ex]
C1_{hA11}&=\Bigg(3 m_B^4 m_C^4 \bigg(m^2 (6 m^8 m_B^2-2 m^6 (m_A^2-m_B^2) (9 m_B^2+m_C^2)\nonumber \\
&\qquad +m^4 (m_A^2-m_B^2)^2 (19 m_B^2+5 m_C^2)-2 (m_A^2-m_B^2)^4 (m_C^2-m_D^2)\nonumber \\
&\qquad -m^2 (m_A^2-m_B^2)^3 (7 m_B^2-m_C^2+2 m_D^2))-2 (m_A^2-m_B^2)\nonumber \\
&\qquad \times (m^2-m_A^2+m_B^2)^3 (2 m^2 (m_B^2+m_C^2)+(m_A^2-m_B^2) (m_C^2-m_D^2)) \nonumber \\
&\qquad \times \mathrm{log}\left[1-\frac{m^2}{m^2-m_A^2+m_B^2}\right]\bigg)\Bigg)/\Big((m_A^2-m_B^2)^3 (m^2-m_A^2+m_B^2)^3\nonumber \\
&\qquad \times (m_B^2-m_C^2)^2 (2 m_B^2+m_C^2) (m_C^2-m_D^2)^2\Big)
\displaybreak[0] \,, \\[1.5ex]
C2_{hA12}&=\Bigg(3 m_B^4 m_C^4 \bigg(\frac{1}{(m^2-m_A^2+m_B^2)^3 m_C^2 (m_C^2-m_D^2)}\nonumber \\
&\qquad \times\Big(2 m^{10} m_B^2 m_C^2 (5 m_C^2-3 m_D^2)+2 (m_A^2-m_B^2)^5 (m_B^2+2 m_C^2) (m_C^2-m_D^2)^2\nonumber \\
&\qquad -2 m^8 (m_A^2-m_B^2) m_C^2 (-m_C^4+m_C^2 m_D^2+2 m_B^2 (8 m_C^2-5 m_D^2))\nonumber \\
&\qquad -2 m^2 (m_A^2-m_B^2)^4 (m_C^2-m_D^2) (9 m_C^4-7 m_C^2 m_D^2+m_B^2 (2 m_C^2-3 m_D^2))\nonumber \\
&\qquad -m^4 (m_A^2-m_B^2)^3 (-25 m_C^6+39 m_C^4 m_D^2-14 m_C^2 m_D^4\nonumber \\
&\qquad +m_B^2 (11 m_C^4-m_C^2 m_D^2-6 m_D^4))+m^6 (m_A^2-m_B^2)^2\nonumber \\
&\qquad \times (-11 m_C^6+15 m_C^4 m_D^2-4 m_C^2 m_D^4+m_B^2 (35 m_C^4-21 m_C^2 m_D^2-2 m_D^4))\Big)\nonumber \\
&\qquad +2 (m_A^2-m_B^2) (2 m^2 (m_B^2+m_C^2)+(m_A^2-m_B^2) (m_C^2-m_D^2)) \nonumber \\
&\qquad \times \mathrm{log}\left[-\frac{(m^2-m_A^2+m_B^2) (m_C^2-m_D^2)}{m^2 m_C^2}\right]\bigg)\Bigg)\nonumber \\
&\qquad /\Big((m_A^2-m_B^2)^3 (m_B^2-m_C^2)^2 (2 m_B^2+m_C^2) (m_C^2-m_D^2)^2\Big)
\displaybreak[0] \,, \\[1.5ex]
C1_{lA1}&=\Bigg(3 m_B^4 \bigg(-(m_A^2-m_B^2)^5 (m_B^2-m_C^2) (3 m_C^4-4 m_C^2 m_D^2+m_D^4)\nonumber \\
& \qquad +2 m^{10} m_B^2 (-2 m_C^4-2 m_C^2 m_D^2+m_D^4)-m^6 (m_A^2-m_B^2)^2\nonumber \\
& \qquad \times (-10 m_C^6-16 m_C^4 m_D^2+7 m_C^2 m_D^4+m_B^2 (28 m_C^4+28 m_C^2 m_D^2-13 m_D^4))\nonumber \\
& \qquad +m^4 (m_A^2-m_B^2)^3 (-10 m_C^6-22 m_C^4 m_D^2+9 m_C^2 m_D^4\nonumber \\
& \qquad +m_B^2 (14 m_C^4+28 m_C^2 m_D^2-11 m_D^4))+2 m^8 (m_A^2-m_B^2)\nonumber \\
& \qquad \times (-2 m_C^6-2 m_C^4 m_D^2+m_C^2 m_D^4+m_B^2 (9 m_C^4+8 m_C^2 m_D^2-4 m_D^4))\nonumber \\
& \qquad +m^2 (m_A^2-m_B^2)^4 (m_B^2 (3 m_C^4-16 m_C^2 m_D^2+5 m_D^4)\nonumber \\
& \qquad -m_C^2 (m_C^4-14 m_C^2 m_D^2+5 m_D^4))+2 (m_A^2-m_B^2) (m^2-m_A^2+m_B^2)^3\nonumber \\
& \qquad \times m_C^4 (2 m^2 (m_B^2+m_C^2)+(m_A^2-m_B^2) (m_C^2-m_D^2)) \nonumber \\
& \qquad \times \mathrm{log}\left[\frac{(1-\frac{m_A^2}{m_B^2}) m_C^2}{m^2-m_A^2+m_B^2}\right]\bigg)\Bigg)/\Big((m_A^2-m_B^2)^3 (m^2-m_A^2+m_B^2)^3\nonumber \\
& \qquad \times (m_B^2-m_C^2)^2 (2 m_B^2+m_C^2) (m_C^2-m_D^2)^2\Big)
\,.
\end{align}
\paragraph{SFSV:}
\begin{align}
C1_f&=\frac{m_B^2}{\left(-m_A^2+m_B^2\right) \left(-m_B^2+m_C^2\right)}
\displaybreak[0] \,, \\[1.5ex]
C1_n&=\frac{2 m_B^2 m_C^2 \left(-m_B^2+m_C^2+2 m_B^2 \mathrm{log}\left[\frac{m_B}{m_C}\right]\right)}{\left(m_A^2-m_B^2\right) \left(m_B^2-m_C^2\right)^2 \left(m_C^2-m_D^2\right)}
\displaybreak[0] \,, \\[1.5ex]
C2_n&=-\Bigg(2 m_B^2 m_C^2 \bigg(m_C^2 (m^2 m_B^2-(m_A^2-m_B^2) (m_C^2-m_D^2))\nonumber \\
& \qquad +m_B^2 (-m_A^2+m_B^2) (m_C^2-m_D^2) \mathrm{log}\left[\frac{m^2 m_B^2}{(m_A^2-m_B^2) (m_C^2-m_D^2)}\right]\bigg)\Bigg)\nonumber \\
& \qquad /\Big((m_A^2-m_B^2)^2 (m_B^2-m_C^2)^2 (m_C^2-m_D^2)^2\Big)
\displaybreak[0] \,, \\[1.5ex]
C1_{hA11}&=\Bigg(m_B^2 m_C^2 \bigg(m^2 (2 m^6 m_B^2 m_C^2-2 (m_A^2-m_B^2)^3 (m_B^2-2 m_C^2) (m_C^2-m_D^2)\nonumber \\
& \qquad +m^2 (m_A^2-m_B^2)^2 (-9 m_C^4+6 m_C^2 m_D^2+m_B^2 (7 m_C^2-4 m_D^2))\nonumber \\
& \qquad -m^4 (m_A^2-m_B^2) (-3 m_C^4+2 m_C^2 m_D^2+m_B^2 (7 m_C^2-2 m_D^2)))\nonumber \\
& \qquad +2 m_B^2 (m_A^2-m_B^2) (m^2-m_A^2+m_B^2)^3 (m_C^2-m_D^2) \nonumber \\
& \qquad \times \mathrm{log}\left[1-\frac{m^2}{m^2-m_A^2+m_B^2}\right]\bigg)\Bigg)/\Big((m_A^2-m_B^2)^2 (m^2-m_A^2+m_B^2)^3\nonumber \\
& \qquad \times (m_B^2-m_C^2)^2 (m_C^2-m_D^2)^2\Big)
\displaybreak[0] \,, \\[1.5ex]
C2_{hA12}&=\Bigg(2 m_B^2 m_C^2 \bigg((m^6 m_C^2 (m_B^2+3 m_C^2-2 m_D^2)\nonumber \\
& \qquad +2 m_B^2 (-m_A^2+m_B^2)^3 (m_C^2-m_D^2)+2 m^2 (m_A^2-m_B^2)^2\nonumber \\
& \qquad \times (3 m_B^2 m_C^2+2 m_C^4-2 (m_B^2+m_C^2) m_D^2)-m^4 (m_A^2-m_B^2)\nonumber \\
& \qquad \times (9 m_C^4-6 m_C^2 m_D^2+m_B^2 (5 m_C^2-2 m_D^2)))/(2 (m^2-m_A^2+m_B^2)^3)\nonumber \\
& \qquad +m_B^2 (m_C^2-m_D^2) \mathrm{log}\left[\frac{m^2 m_C^2}{(m^2-m_A^2+m_B^2) (-m_C^2+m_D^2)}\right]\bigg)\Bigg)\nonumber \\
& \qquad /\Big((m_A^2-m_B^2) (m_B^2-m_C^2)^2 (m_C^2-m_D^2)^2\Big)
\displaybreak[0] \,, \\[1.5ex]
C1_{lA1}&=\Bigg(2 m_B^2 m_C^2 \bigg(\Big(-2 m^8 m_B^2 m_C^4+(m_A^2-m_B^2)^4 (m_B^2-m_C^2) (m_C^4-m_D^4)\nonumber \\
& \qquad -m^2 (m_A^2-m_B^2)^3 (m_C^2-m_D^2) (m_C^4-3 m_C^2 m_D^2+m_B^2 (m_C^2+3 m_D^2))\nonumber \\
& \qquad +m^6 (m_A^2-m_B^2) (-m_C^2 (2 m_C^4-2 m_C^2 m_D^2+m_D^4)\nonumber \\
& \qquad +m_B^2 (6 m_C^4-2 m_C^2 m_D^2+m_D^4))-m^4 (m_A^2-m_B^2)^2\nonumber \\
& \qquad \times (-3 m_C^2 (2 m_C^4-2 m_C^2 m_D^2+m_D^4)+m_B^2 (4 m_C^4-4 m_C^2 m_D^2+3 m_D^4))\Big)\nonumber \\
& \qquad /\Big(2 (m_A^2-m_B^2)(m^2-m_A^2+m_B^2)^3 m_C^2\Big)+m_B^2 (-m_C^2+m_D^2)\nonumber \\
& \qquad \mathrm{log}\left[\frac{(m_B^2-m_A^2) m_C^2}{m_B^2(m^2-m_A^2+m_B^2)}\right]\bigg)\Bigg)/\Big((m_A^2-m_B^2) (m_B^2-m_C^2)^2 (m_C^2-m_D^2)^2\Big)
\,.
\end{align}


\end{document}